\documentclass[twocolumn]{aastex631}

\usepackage{amsmath}

\renewcommand{\d}{\mathrm{d}}

\usepackage{color}
\usepackage{amsmath}

\begin{document}

\title[Radio emission from TDEs]{Radio emission from tidal disruption events produced by the collision between super-Eddington outflows and the circumnuclear medium}

\author[0000-0001-7984-9477]{Fangyi (Fitz) Hu}
\affiliation{School of Physics and Astronomy, Monash University, Clayton VIC 3800, Australia}
\affiliation{OzGrav: The ARC Centre of Excellence for Gravitational Wave Discovery, Australia}

\author[0000-0003-3441-8299]{Adelle Goodwin}
\affiliation{International Centre for Radio Astronomy Research, Curtin University, Perth WA 6845, Australia}

\author[0000-0002-4716-4235]{Daniel J. Price}
\affiliation{School of Physics and Astronomy, Monash University, Clayton VIC 3800, Australia}

\author[0000-0002-6134-8946]{Ilya Mandel}
\affiliation{School of Physics and Astronomy, Monash University, Clayton VIC 3800, Australia}
\affiliation{OzGrav: The ARC Centre of Excellence for Gravitational Wave Discovery, Australia}

\author[0000-0002-1084-3656]{Re'em Sari}
\affiliation{Racah Institute of Physics, The Hebrew University of Jerusalem, 9190401, Israel}

\author[0000-0003-4799-1895]{Kimitake Hayasaki}
\affiliation{Department of Astronomy and Space Science, Chungbuk National University, Cheongju 361-763, Republic of Korea}
\affiliation{Department of Physical Sciences, Aoyama Gakuin University, Sagamihara 252-5258, Japan}



\begin{abstract}

In this Letter, we simulate the collision between outflows from the tidal disruption of a 1M$_\odot$ main sequence star around a $10^6$M$_\odot$ black hole and an initially spherically symmetric circumnuclear cloud. We launch super-Eddington outflows self-consistently by simulating the disruption of stars on both bound and unbound initial orbits using general relativistic smoothed particle hydrodynamics. We find shocks formed as early as $\sim 10~$days after the initial stellar disruption produce prompt radio emission. The shock radius ($\approx~10^{17}$~cm), velocity ($\sim 0.15$c) and total energy ($\sim 10^{51}$ erg) in our simulations match those inferred from radio observations of tidal disruption events (TDEs).
We ray-trace to produce synthetic radio images and spectra to compare with the observations.  
While the TDE outflow is quasi-spherical, the synchrotron emitting region is aspherical but with reflection symmetry above and below the initial orbital plane. Our synthetic spectra show continuous decay in peak frequency, matching prompt radio TDE observations.
Our model supports the hypothesis that synchrotron radio flares from TDEs result from the collision between outflows and the circumnuclear material.

\end{abstract}



\section{Introduction} \label{sec:intro}

Stars approaching supermassive black holes (SMBHs) are disrupted by the strong tidal force, resulting in tidal disruption events (TDEs). The expected outcome from the disruption is that mass falling back towards the black hole rapidly circularises to form an accretion disk, resulting in soft X-ray emission \citep{rees1988}.

While some TDE candidates are observed in X-rays \citep[e.g.][]{Auchettl2017, Gezari2017, Jonker2020, Chen2022a, Wevers2024}, most have been discovered in optical/UV bands \citep[e.g.][]{van-Velzen2011,van-Velzen2021a,Hammerstein2023,Yao2023a} associated with strong outflows of $\sim 10^{4}$ km/s, likely as a result of reprocessing in an outflowing, optically thick envelope of material \citep[e.g.][]{Loeb1997, Strubbe2009a, Piran2015, Parkinson2022a}, as demonstrated in recent simulations \citep[e.g.][]{Hu2024,Price2024a}. Even more surprising was the discovery of radio emission associated with TDE flares. Both prompt \citep[within weeks after the optical peak; e.g.][]{Alexander2016, Cendes2021, Goodwin2022, Goodwin2023, Goodwin2023a, Goodwin2024, Dykaar2024a} and delayed \citep[months to years after the optical peak; e.g.][]{Horesh2021a, Horesh2021b, Perlman2022, Sfaradi2022, Cendes2024a, Cendes2024} radio emission is observed from TDEs, raising questions about its origin.

Two hypotheses for the origin of synchrotron radio flares from TDEs are: 1) internal shocks of mildly collimated, sub-relativistic TDE jets \citep[e.g.][]{van-Velzen2011b,van-Velzen2016b,Pasham2018}, and 2) collision (external shocks) between the TDE outflow (accretion-induced, self-collision-induced, or unbound debris stream) or jets and circumnuclear material (CNM) \citep[e.g.][]{Alexander2016,Krolik2016, Xu2022,Goodwin2023}. Several analytical models try to explain the prompt \citep[e.g.][]{Metzger2012, Krolik2016, Hayasaki2023} and late \citep[e.g.][]{Xu2022, Matsumoto2024a, Zhuang2024} radio emissions with the second hypothesis, however, to date it has not been explored with hydrodynamic simulations. 

\citet{Goodwin2022} extrapolated the radius of the emitting region backward in time and showed that the launch time of the outflow producing prompt radio emission is coincident with the beginning of the optical flare. This shows that the prompt radio emissions could originate from the same outflow as the optical emission.

%
This hypothesis of radio emission from outflows colliding with the CNM is also supported by the recent IceCube detected high-energy neutrino sources that are spatially and temporally coincident with TDEs or TDE candidates \citep{Stein2021, Stein2023, Jiang2023, Yuan2024b}. The shocks between TDE outflows and the CNM are capable of producing the observed neutrinos \citep[e.g.][]{Murase2020, Wu2022, Piran2023}, although such a model would predict $\gamma$-rays that are yet to be observed \citep[e.g][]{Chen2016, Mou2021, Peng2022}. 

In our previous paper \citep{Hu2024} we simulated the tidal disruption of a $1 M_{\odot}$ star on an eccentric orbit ($e=0.95$) around a $10^6 M_{\odot}$ black hole. Similar to the study by \citet{Price2024a}, we showed that the tidal disruption of the star and subsequent accretion onto the black hole produced quasi-spherical super-Eddington outflows that could explain the otherwise-mysterious optical emission associated with observed TDE candidates.

In this Letter, we simulate the collision between these TDE outflows and the CNM, study the shock regions by
comparing to the observational inferred properties,
and predict the synchrotron emission,
to test whether this model can explain the observed radio emission. We describe the setup and methods in Section~\ref{sec:methods} and show the results and analysis in Section~\ref{sec:results}. We finally discuss the implications and future improvements in Section~\ref{sec:discussion}.

\section{Methods} \label{sec:methods}
We use the general relativistic smoothed particle hydrodynamics code \textsc{Phantom} \citep{Price2018,Liptai2019a} for the simulations, assuming a fixed Schwarzschild background metric with a $10^6 M_{\odot}$ central black hole. 

To initiate our experiments, we used the output from two previous simulations of eccentric TDEs ($e=0.95$) taken from \citet{Hu2024} with penetration factors of $\beta=1$ and 5.
The penetration factor is defined as 
\begin{equation}
    \beta \equiv r_{\rm t}/r_{\rm p},
\end{equation}
where $r_{\rm t} = R_* (M_{\rm BH}/M_*)^{1/3}$ is the tidal radius and $r_{\rm p}$ is the pericenter distance. The resolution of the simulation with $\beta = 5$ is $10^6$ SPH particles and the one with $\beta = 1$ is $10^5$. The stellar model and setup is described in \citet{Hu2024} but was mapped into 3D from a 1D stellar model evolved to 4.6 Gyr in the {\sc mesa} stellar evolution code. We refer to these four simulations as `TDE simulations'.

To simulate the interaction with the CNM, we continuously inject the outflowing material from each of the two TDE simulations into a new simulation that contains an initially static, spherically symmetric CNM shell (Section~\ref{sec:cnm setup}). We refer to this set of simulations as `radio simulations'. These simulations are also performed in the Schwarzshild metric. 

Accretion onto the black hole in the TDE simulations is initiated by self-collision of the stream caused by general relativistic apsidal precession, as already found by \citet{Bonnerot2016a}, \citet{Hayasaki2016} and subsequent authors. In our TDE simulations, we assume an adiabatic equation of state (EoS) which assumes the gas is optically thick. The same assumption is made by \citet[][e.g. their Figure 5]{Bonnerot2016a}. { The adiabatic approximation assumes that all radiation released is trapped by gas and that gas and radiation temperature are in equilibrium. Importantly, the assumption of equal gas and radiation temperatures is only valid inside the optical photosphere and hence only for our `TDE simulations' but not for our `radio simulations'}. The trapping of radiation increases the pressure and naturally creates outflows powered by heat generated from material falling close to the black hole. Such outflows do not occur in isothermal or isentropic simulations where the accretion energy is removed from the simulation \citep{Bonnerot2016a,Price2024a}. To capture intermediate behaviour one would need to incorporate radiation hydrodynamics. However, the adiabatic approximation is likely closer to reality given the high column density and hence high optical depth of the accreting material.




{ For the radio simulations, }we inject outflowing material from the original simulations once it passes a certain radius (Section~\ref{sec:injection}) to study the subsequent interactions. The simulations are evolved until three years after the first pericenter passage. We then post-process each simulation to analyse the shock region formed, estimate properties of the shock region (Section~\ref{sec:shock properties}) to compare with properties inferred from an equipartition model (Section~\ref{sec:equipartition model}) in observations, and ray-trace the synchrotron emissions to create synthetic images and spectra (Section~\ref{sec:synchrotron spectra}). 

\begin{figure*}
 \centering
    \includegraphics[width=2\columnwidth]{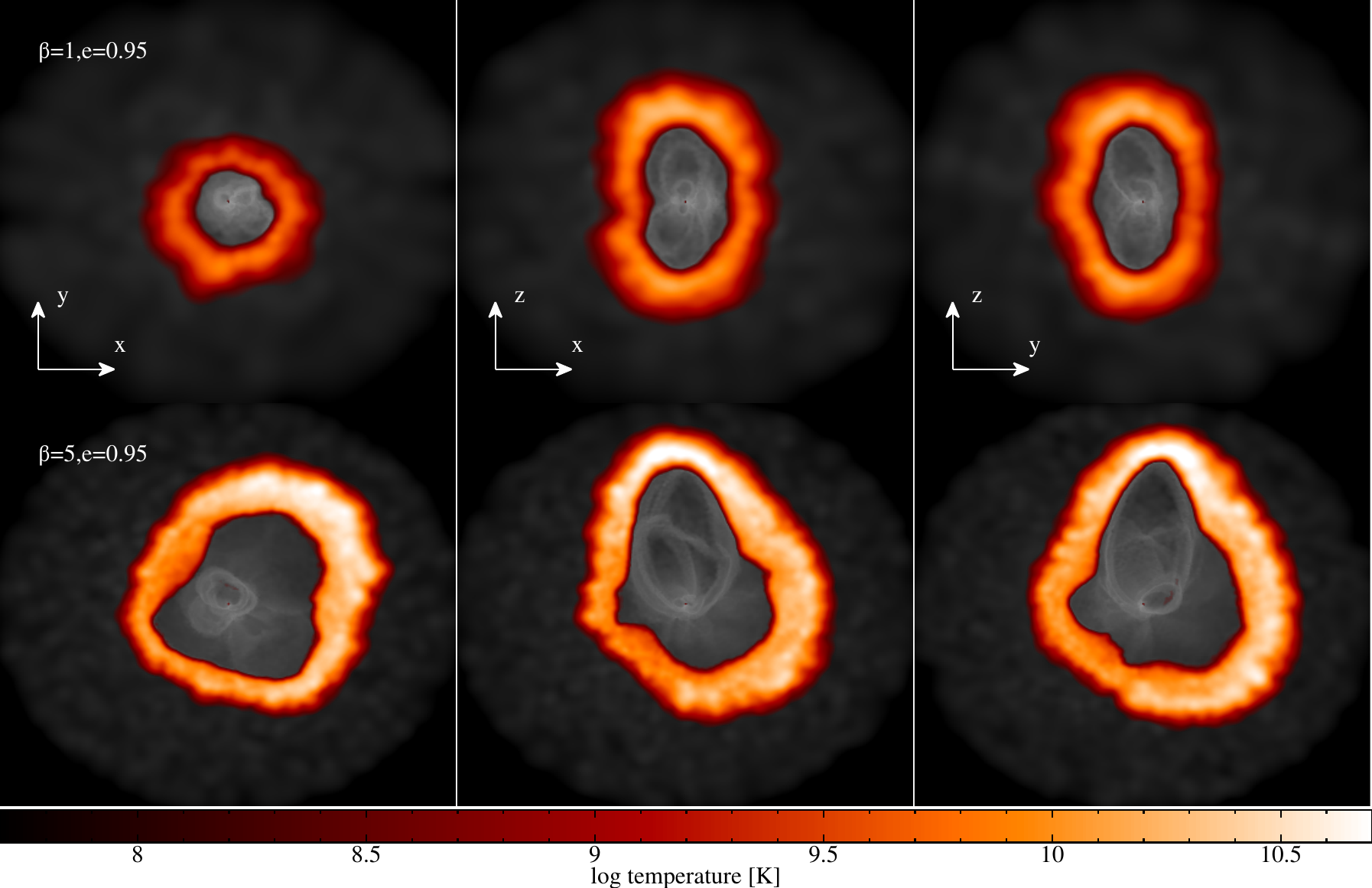}
    \caption{x-y (left column), x-z (middle column) and y-z plane (right column) 0.4 pc $\times$ 0.4 pc cross-sections of the outflowing shock shells between various TDEs (top row: $\beta=1,e=0.95$; bottom row: $\beta=5,e=0.95$) and CNM cloud of 0.1 $M_{\odot}$ at $t=3~$yr post first pericenter passage, rendered with gas temperature (coloured; pixels with $T$ below lower limit are transparent) and density (grayscale; black is $10^{-23}$ g cm$^{-3}$, white is $8.3\times 10^{-13}$ g cm$^{-3}$). The initial TDE stellar orbit is anti-clockwise in the x--y plane and the pericenter is in the $-x$ direction. Outflows are launched from near the BH (center of each panel; { $r_{\rm inj}=3 \times 10^{-4}$pc is smaller than one pixel in the image}) and interact with surrounding CNM, which forms a hot shock shell (coloured region). We assumed a constant electron energy fraction $\epsilon_{\rm e}$, so the gas temperature represents the electron temperature. Simulation files and animations are available on Zenodo: doi: 10.5281/zenodo.14286338. The animation shows the time evolution.}
    \label{fig:shock_view}
\end{figure*}

\subsection{Circumnuclear material} \label{sec:cnm setup}
In the radio simulations, we set up the spherically symmetric CNM clouds between an inner radius of $r_{0} = 10^{15}~$cm, where within this radius we assume there is insignificant amount of CNM, and an outer radius of $r_{1} = 6\times10^{17}~$cm, which is roughly the furthest distance outflowing material in the TDE simulations would reach by the end of three years. We assume a power law density profile, i.e.
\begin{equation}
    \rho = \rho_{0} (r/r_{0})^{n},
    \label{eq:rho}
\end{equation}
in the CNM. To ensure significant collisions between TDE outflow and CNM cloud, we set the total mass of the CNM cloud to be $0.1M_{\odot}$, i.e. comparable to the outflow mass. The normalisation factor $\rho_{0}$ required to reach the total mass $M_{\rm CNM}$ can be calculated from
\begin{equation}
    \rho_{0} = \frac{(3+n)M_{\rm CNM}r_{0}^{n}}{4\pi\left(r_{1}^{3+n} - r_{0}^{3+n} \right)},
\end{equation}
where $n$ is the density power law index. In this work, we adopt $n=-1.7$ from AT2019dsg \citep{Cendes2021}, which gives $\rho_0= 5.03\times10^{-18}~$g cm$^{-3}$.

The CNM shell is assumed to be initially stationary, i.e. $v=0$, and cold, i.e. $T = 10$ K. We use a gas-only adiabatic equation of state (EoS), i.e. $P = (\gamma-1)\rho u$, where $\gamma = 5/3$ and internal energy is related to temperature by
\begin{equation} \label{eq:gastemp}
    u = \frac{3}{2}\frac{k_{\rm B}T}{\mu m_{\rm H}}.
\end{equation}
{ This is valid since our `radio simulations' are performed outside the optical photosphere where the gas and radiation temperatures are no longer equal.}
We calculate the specific entropy of each particle according to the Sackur-Tetrode equation,
\begin{equation}
s = \frac{k_{\rm B}}{\mu m_{\rm H}} \ln\left({\frac{T^{3/2}}{\rho}}\right), \label{eq:s}
\end{equation}
with $\mu = 0.5988$. $s$ is also the energy variable we evolve in the simulations (see \citealt{Liptai2019a}). The entropy per baryon is $S/k_{\rm B} = sm_{\rm H}/k_{\rm B}$.


\subsection{Collisionless shocks}
In the environment of TDE outflows and CNM, the gas density is too low that collisional shock formation is not possible since the mean free path of the gas particles exceeds the shock width. However, collisionless shocks can still form due to the presence of charged particles in the ambient magnetic field \citep[e.g.][]{Inoue2011, Tomita2022}. Inside the optical photosphere the outflow is expected to be radiation-pressure dominated, with $\gamma = 4/3$. This assumption explains the low $\sim 10^4$K temperature observed at the optical photosphere { in the TDE simulations}, since $\rho u \approx a T^4$. However, in optically thin regions outside the photosphere { (as in our radio simulations)} the gas is, by definition, no longer in equilibrium with the radiation field, so the appropriate temperature is the gas temperature computed from Eq.~\eqref{eq:gastemp}. This produces typical temperatures of $10^{10}$--$10^{11}$~K { as shown in Figure~{\ref{fig:shock_view}}}. However, both of these temperatures are merely interpretations of the internal energy in the respective calculations, and neither temperature is assumed when computing the synchrotron spectrum which is computed directly from the available energy in the shock { in the radio simulations. We emphasise that the abrupt change in temperature seen in Figure~\ref{fig:shock_view} is caused by the increased internal energy behind the shock, not any change in the temperature prescription. That is, we use Equation~\eqref{eq:gastemp} everywhere when converting internal energy to temperature in the radio simulations, not just at the shock location.}

\subsection{Outflow injection} \label{sec:injection}
TDE outflows enter the radio simulation from within the inner radius. 
We chose to continuously inject outflow by comparing between TDE simulation outputs saved every 4.38 hours, giving 2000 snapshots for each 365-day TDE simulation. The injection radius $r_{\rm inj}$ is chosen to be $9.8\times 10^{14}~$cm to avoid direct contact between the injected outflow and CNM shell. 

To determine if a particle $i$ will be injected, we compare its positions between TDE simulation snapshots. If its radius in the current snapshot $r_{i}$ is greater than $r_{\rm inj}$ and its radius in the previous dump $r_{i,{\rm pre}}$ is smaller than $r_{\rm inj}$, we decide this particle is newly ejected and we inject it into the radio simulations. An `injected' status flag is given to each injected particle so they will be ignored in the future injections even if they cross the injection radius again in future in the TDE simulation. 

After the interaction with the CNM, the particle trajectory will differ from the TDE simulation. The particles could keep outflowing or fall towards the BH. If a particle falls through the injection radius, i.e. $9.8\times 10^{14}~$cm, we remove the particle for the rest of the radio simulation to save computational resources.

\subsection{Shock properties} \label{sec:shock properties}

We assume that synchrotron emission arises from non-thermal acceleration of electrons in the shocked region between the outflow and the CNM. Since only material in the shocked region would accelerate particles to relativistic speeds, we first identify shocked material in our simulation, in order to identify the amount of energy available for particle acceleration.
To determine if an SPH particle has been shocked, we compare its entropy (Eq.~\ref{eq:s}) against the background value, i.e. value in the initial conditions or when injected. If the entropy rises above the background value by more than 30\%, the particle is considered to be in the shocked region and emitting synchrotron radiation. The choice of the threshold is not critical within 10\% to 60\% (see Section~\ref{sec:shock profile}).




To compare with observations \citep[e.g.][]{Goodwin2023}, we estimated the radius, velocity and total energy of the shocked region, which can all be inferred directly from our radio simulations. 

With all the shocked material determined, we estimated the shock properties solely from this material. 
{ The TDE outflows in our simulations have a non-spherical nature, which can be seen as a radially extended shocked gas layer when projected as a function of spherical radius in Figure~\ref{fig:shock_profile}. We therefore take} the maximum radius and maximum velocity of all shocked particles (i.e. along all lines of sight) as the shock radius and velocity respectively { when comparing to the observations. This gives an upper limit for the spherically-averaged shock radius and velocity. A better procedure would be to produce synthetic spectra from first principle (see Section~\ref{sec:synchrotron spectra}) and compare directly to the observations or with same assumed model.}

The total shock energy is considered as the total energy of all SPH particles in the shocked region, minus the rest mass energy. In GR this corresponds to \citep[e.g.][]{Monaghan2001}
\begin{equation}
E_{\rm shock} = \sum_{a=1}^{N} m_{a} \left[p_{i,a} v^{i}_a + \frac{1}{U_a^0} \left( c^2 + u_a \right)\right] - \sum_{a=1}^N m_a c^2,
\end{equation}
for $N$ particles with mass $m$, where $p_i$ is the specific momentum, $v^i$ is the velocity, $u$ is the specific thermal energy and $U^0$ is the time component of the four velocity.
At sub-relativistic speeds far away from the black hole this becomes simply
\begin{equation}
E_{\rm shock} = \sum_{a=1}^{N} m_{a} \left[\frac12 v_{a}^{2} + u_a - \frac{GM_{\rm BH}}{r_a} \right],
\end{equation}
with radius $r$. We account for the gravitational potential energy here for completeness, but it is $\gtrsim 5$ orders of magnitude smaller than the kinetic and internal energy at all time. We define the specific total energy per SPH particle in the shock as 
\begin{equation}
e_{\rm shock} \equiv \frac12 v^{2} + u - GM_{\rm BH}/{r}. \label{eq:eshock}
\end{equation}




\begin{table*}[]
    \centering
    \begin{tabular}{c|ccccc}
         TDE & $f_{A}$ & $f_{V}$ & $\epsilon_{\rm e}$ & $\epsilon_{\rm B}$ & Source\\
        \hline
        ASAASN-14li & 1, 0.1 & 0.36 & 0.1 & 0.5 & \citet{Alexander2016} \\
        CNSS J0019 & 1 & 0.5 &  1/3 & 1/3 & \citet{Anderson2020} \\
        AT2019dsg & 1 & 0.36 & 0.1 & 0.02 & \citet{Cendes2021}\\
        AT2019azh & 1, 0.13 & 4/3, 1.15 & 0.1 & $10^{-3}$ & \citet{Goodwin2022} \\
        AT2020opy & 1, 0.13 & 4/3, 1.15 & 0.1 & $10^{-3}$ & \citet{Goodwin2023a} \\
        AT2020vwl & 1, 0.13 & 4/3, 1.15 & 0.1, $5\times 10^{-4}$ & 0.02 & \citet{Goodwin2023} \\
        eRASSt J2344 & 1, 0.13 & 4/3, 1.15 & 0.1, $10^{-3}$, $10^{-4}$ & 0.02 & \citet{Goodwin2024}
    \end{tabular}
    \caption{Outflow geometry (area filling factor $f_{\rm A}$, volume filling factor $f_{\rm V}$) and energy partition (electron energy fraction $\epsilon_{\rm e}$, magnetic energy fraction $\epsilon_{\rm B}$) assumptions used in each analysis. Note that \citet{Anderson2020} used the equipartition model from \citet{Chevalier1998} whereas others used the \citet{Barniol-Duran2013} model.}
    \label{tab:obs}
\end{table*}

\subsection{Equipartition model} \label{sec:equipartition model}
In radio observations of TDEs, the equipartition model of synchrotron emission is extensively used to infer physical outflow properties from observed synchrotron spectral peak flux densities and frequencies \citep{Barniol-Duran2013}. It utilises the opposite dependencies of magnetic and particle energy density on the magnetic field strength, i.e. $U_{\rm B} \propto B^2$ and $U_{\rm E} \propto B^{-3/2}$, respectively. There is, therefore, a minimum total energy density near equipartition, i.e. at $U_{\rm E} \approx U_{\rm B}$, and a lower limit of the energy can be estimated from the observations with this model.

By fitting to the synchrotron spectra, a power law index $p$ of the electron energy distribution can be obtained, i.e. $N(E) \propto E^{-p}$. With additional assumptions on the geometry of the emitting region, namely an area filling factor $f_{\rm A} = A/(\pi R^2/\gamma^2)$ and a volume filling factor $f_{\rm V} = V/(\pi R^3/\gamma^4)$, where $\gamma$ is the bulk Lorentz factor of the emitting region, and the fraction of total energy carried by electrons, $\epsilon_{\rm e}$, the equipartition energy $E_{\rm eq}$ and radius $R_{\rm eq}$ can be estimated, as well as the bulk velocity $V$, magnetic field strength $B$ and electron density $n_{\rm e}$ of the emitting region. The deviation from the equipartition can be parametrised by both $\epsilon_{\rm e}$ and the fraction of total energy in the magnetic field, $\epsilon_{\rm B}$. The detailed equations for all the aforementioned properties can be found in \citet{Barniol-Duran2013}. 

In the case of TDEs, observers refer to the conical geometry with $f_{\rm A} \lesssim 0.1$ as a jet and a quasi-spherical geometry with $f_{\rm A} \to 1$ as an outflow. These two geometries are extensively used \citep[e.g.][]{Alexander2016, Goodwin2022, Goodwin2023, Goodwin2023a, Goodwin2024} to account for the uncertainties in the sources of radio observations, e.g. the two hypotheses mentioned in Section~\ref{sec:intro}. 

\subsection{Synchrotron spectra} \label{sec:synchrotron spectra}
In addition to direct comparisons with the equipartition model, we also tried to predict the synchrotron emission from our simulations self-consistently.

We used a similar ray tracing method as \citet{Hu2024} and \citet{Price2024a} except with source function and opacity for synchrotron emission. From the simulations, we assume only particles in the shocked region (determined as outlined in Section \ref{sec:shock properties}) will be emitting synchrotron emission and each SPH particle as an individual emission region with the same bulk properties. 

We assume that the electrons are accelerated in the shock into a power law distribution with a minimum energy $E_{\rm m}$, i.e.
\begin{equation} \label{eq:e plaw}
    N(E_{\rm e}) = N_{0} E^{-p}_{\rm e}\ {\rm for}\ E_{\rm e} > E_{\rm m},
\end{equation}
where $N_{0} = (p-1) N_{\rm e} E_{\rm m}^{p-1}$ with number of free electrons $N_{\rm e}$. The energy of each electron is $E_{\rm e} = (\gamma_e - 1)m_ec^2$ where $\gamma_{\rm e}$ is the Lorentz factor of each electron and $m_{\rm e}$ and $c$ are the electron mass and speed of light respectively. Since the shocked region has temperature $\gtrsim 10^8$~K (Figure~\ref{fig:shock_profile}), we assume the gas is fully ionised and pure hydrogen, which gives $N_{\rm e} = N_{\rm p} = m_{\rm sph}/m_{\rm p}$ with $m_{\rm sph}$ and $m_{\rm p}$ being the mass of each SPH particle and the proton mass respectively. 

\subsubsection{Source function} \label{sec:source function}

To get the source function of synchrotron emission, we neglect scattering and consider only emission and absorption \citep{Rybicki1979}. Then
\begin{equation} \label{eq:source func}
    S_{\nu} = j_{\nu}/\alpha_{\nu},
\end{equation}
where the emission coefficient is
\begin{equation} \label{eq:emi coe}
    j_{\nu} = P_{\nu}/4\pi,
\end{equation}
assuming isotropic radiation with power $P_{\nu}$, and the absorption coefficient is \citep{Rybicki1979}
\begin{align} \label{eq:abs coe}
    \alpha_{\nu} = \frac{\sqrt{3}q_{\rm e}^3}{8\pi m_{\rm e}} &\left(\frac{3q_{\rm e}}{2\pi m_{\rm e}^3 c^5}\right)^{\frac{p}{2}}N_{0}(B\sin{\alpha})^{\frac{p+2}{2}} \nonumber \\
    &\Gamma\left(\frac{3p+2}{12}\right)\Gamma\left(\frac{3p+22}{12}\right)\nu^{-\frac{p+4}{2}}, 
\end{align}
where $q_{\rm e}$ is the electron charge and $B, \alpha$ and $\nu$ are the magnetic field, pitch angle and radiating frequency respectively. $N_{0}$ and $p$ are the parameters of the assumed power distribution of electrons from Equation \eqref{eq:e plaw}, and $\Gamma$ is the gamma function.

We follow \citet{Sari1998} to calculate $P_{\nu}$ for each particle, except that we do not need to make assumptions about the the jump conditions \citep{Blandford1976}. Instead, we use the post-shock fluid properties directly (Section~\ref{sec:shock properties}) and simply compute the fraction of post-shock energy contained by electrons and magnetic field, $\epsilon_e$ and $\epsilon_B$, respectively, according to
\begin{align}
    \int^\infty_{E_{\rm m}} E_e N(E_e) dE_e & = \epsilon_e e_{\rm shock}, \\
   \frac{B^2}{8\pi \rho} & = \epsilon_{B} e_{\rm shock},
\end{align}
giving corresponding expressions for the magnetic field and the minimum frequency $\nu_{\rm m}$
\begin{align}
    B &= \left( 8 \pi \epsilon_{B} e_{\rm shock} \rho \right)^{1/2}, \label{eq:mag strength} \\
    \nu_{\rm m} &= \frac{\gamma q_{e}B}{2\pi m_{\rm e}c}\left(\frac{\epsilon_{\rm e}m_{\rm p}e_{\rm shock}}{m_{\rm e}c^2}\frac{p-2}{p-1} + 1\right)^2 \label{eq:nu m},
\end{align}
where we assume cgs units, and $e_{\rm shock}$ and $\gamma$ are the total specific energy (Eq.~\ref{eq:eshock}) and bulk Lorentz factor of each SPH particle, respectively.

We calculate the cooling frequency $\nu_{\rm c}$ in the same way as \citet{Sari1998}, i.e.
\begin{align}
    \nu_{\rm c} &= \frac{18 \pi q_{\rm e} m_{\rm e} c }{\sigma_{\rm T}^{2}} \gamma^{-1} B^{-3}t^{-2} , \label{eq:nu c}
\end{align}
where $\sigma_{\rm T}$ is the Thomson cross section, and $t$ is the cooling time which is assumed to be the interval between the time of interest and the time when the particle is initially shocked. 

\citet{Sari1998} give two regimes for synchrotron emission, namely slow cooling when $\nu_{\rm m} < \nu_{\rm c}$ and fast cooling when $\nu_{\rm c} < \nu_{\rm m}$. We only consider slow cooling in our calculations since for our sub-relativistic outflow $\nu_{\rm m}$ is lower than $\nu_{\rm c}$ for a long time before being cooled. For slow cooling, the spectrum is
\begin{equation} \label{eq:spec}
    P_{\nu} = \left\{
    \begin{array}{ll}
        (\nu/\nu_{\rm m})^{\frac{1}{3}} P_{\nu, {\rm max}} & \nu_{\rm m} \geq \nu \\
        (\nu/\nu_{\rm m})^{-\frac{p-1}{2}} P_{\nu, {\rm max}} & \nu_{\rm c} \geq \nu > \nu_{\rm m} \\
        (\nu_{\rm c}/\nu_{\rm m})^{-\frac{p-1}{2}}(\nu/\nu_{\rm c})^{-\frac{p}{2}} P_{\nu, {\rm max}} & \nu > \nu_{\rm c}
    \end{array}
    \right.,
\end{equation}
where $P_{\nu, {\rm max}} = n_{\rm e} P_{\nu, {\rm e,max}} = n_{\rm e}\gamma B m_{\rm e}c^2\sigma_{\rm T}/3q_{\rm e}$ is the maximum emitting power with electron density $n_{\rm e}$. Since $\nu_{\rm c}$ decreases with time whereas $\nu_{\rm m}$ remains relatively constant, eventually $\nu_{\rm c}$ drops below $\nu_{\rm m}$ and all the electrons within the particle are considered cooled and no future emission is considered from this particle.

Combining Equations \eqref{eq:source func}, \eqref{eq:emi coe}, \eqref{eq:abs coe} and \eqref{eq:spec}, we obtain a source function for synchrotron emission. The shocked region is optically thin, with a downstream temperature high enough to emit X-rays, enabling radiative cooling through free-free emission (Bremsstrahlung cooling). However, since the density in the shocked region is too low to significantly affect its dynamics, we do not consider Bremsstrahlung cooling in our simulations.

\subsubsection{Ray tracing} \label{sec:ray tracing}
Similar to \citet{Hu2024}, assuming the source function $S_{\nu,i}$ and optical depth $\tau_{\nu,i}$ are constants within the region of an individual SPH particle $i$, we solve the radiative transfer equation
\begin{equation}
    I_{\nu,i} = I_{\nu, i-1} e^{-\tau_{\nu,i}} + S_{\nu, i} \left( 1 - e^{-\tau_{\nu,i}} \right), \label{eq:raytrace}
\end{equation}
where $S_{\nu}$ is the source function (Section \ref{sec:source function}) and the frequency-dependent optical depth through each SPH particle (instead of a grey one as in \citealt{Hu2024}) is
\begin{equation} \label{eq:optical depth}
    \tau_{\nu,i} =  \kappa_{\nu,i} m_{i} Y(|x_{{\rm pix},j} - x_i|,|y_{{\rm pix},j} - y_i|, h_i),
\end{equation}
where $m_i$ and $h_i$ are the mass and smoothing length of particle $i$ respectively, $\kappa_{\nu,i} = \alpha_{\nu}/\rho_{i}$ is the opacity at frequency $\nu$, $Y$ is the column-integrated kernel function with dimensions of inverse area (i.e. the 3D spherical kernel function integrated through z, see Eq. 29 of \citealt{Price2007}) and $x_{{\rm pix},j}$ and $y_{{\rm pix},j}$ are the x- and y- coordinates of the image pixel $j$. Using Equations~(\ref{eq:optical depth}) along with (\ref{eq:source func})--(\ref{eq:spec}) we solve (\ref{eq:raytrace}) by first sorting the SPH particles along the line of sight and then summing Eq.~(\ref{eq:raytrace}) from back to front along each line of sight on a grid of $1024 \times 1024$ pixels corresponding to the observational plane.

\subsubsection{Parameter space}
The three key parameters for synchrotron emission, i.e. $\epsilon_{\rm e}$, $\epsilon_{\rm B}$ and $p$, are not well constrained from the TDE observations. For every radio simulation, we performed a grid of synthetic observations with the standard $p=2.5$ \citep{Sari1996} and $\epsilon_{\rm e}$ and $\epsilon_{\rm B}$ each set to 0.5, 0.1 and 0.01. 

\subsubsection{Temporal evolution}
We study the temporal evolution of our synthetic spectra by comparing the peak frequency evolution with the observations. The peak frequency is where the peak flux occur for each of our synthetic spectra as well as the observed spectra. We then fitted a power law to each of the models and observed TDEs to compare their evolution.


\begin{figure}
        \centering
        \includegraphics[width=0.99\columnwidth]{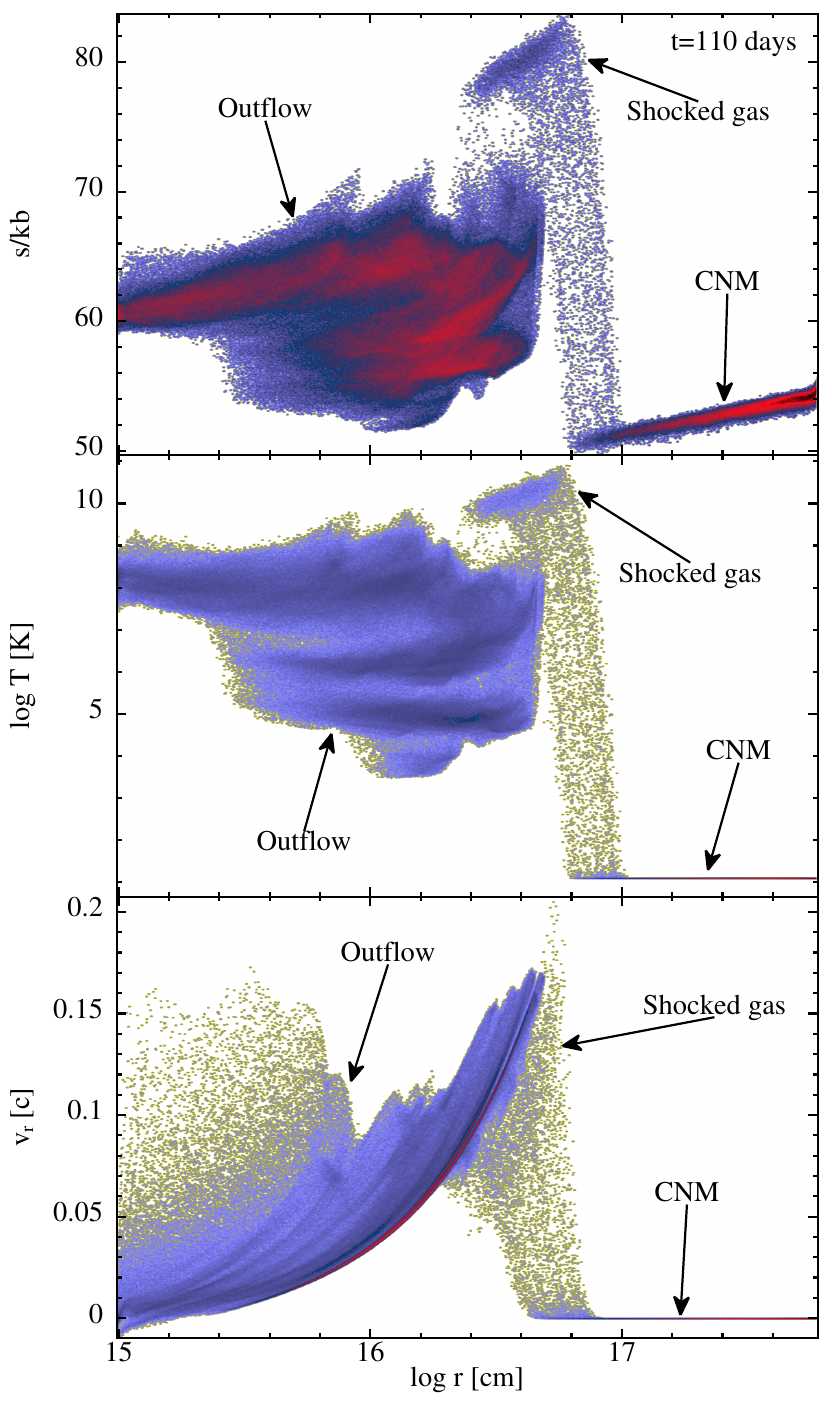}
    \caption{Entropy per baryon (top panel), temperature (middle panel) and radial velocity (bottom panel) as a function of radius in the radio simulation with initial $\beta=5$ and $M_{\rm CNM} = 0.1 M_{\odot}$ at $t=0.3$ yr. A strong forward shock with upstream Mach number $\gtrsim 10^4$ forms between the TDE outflow and the CNM, raising the entropy by $\sim 60\%$ and temperature to $\sim 10^{10}$ K. Simulations files are available on Zenodo: doi: 10.5281/zenodo.14286338. Animations of the evolution of the density, entropy per baryon, temperature and radio velocity as a function of radius for all of our simulations are available.}
    \label{fig:shock_profile}

\end{figure}

\section{Results} \label{sec:results}
\subsection{Temperature images}

Figure~\ref{fig:shock_view} shows the temperature cross sections in the x--y (left), x--z (middle) and y--z (right) planes at three years post-first pericenter passage for the four simulations used. In all the plots, the initial TDE stellar orbit is anti-clockwise in the x--y plane, i.e. angular momentum in $+z$ direction, and the pericenter is in the $-x$ direction.

A quasi-ellipsoidal shock wave is formed at the interface of TDE outflow and CNM with the elongated axis along the z-axis. The TDEs produce outflows collimated along the direction of initial stellar orbital angular momentum 
since all of the angular momentum is deposited into the outflow.

In the simulation with $\beta = 1$, the outflow is quasi-symmetric to the initial stellar orbital plane, i.e. x-y plane. This is expected since there is no initial linear momentum in the z-direction. It is unclear if the asymmetric outburst in the $\beta = 5$ one is physical or numerical. More detailed investigations are necessary to understand this.

\begin{figure}
        \centering
        \includegraphics[width=0.99\columnwidth]{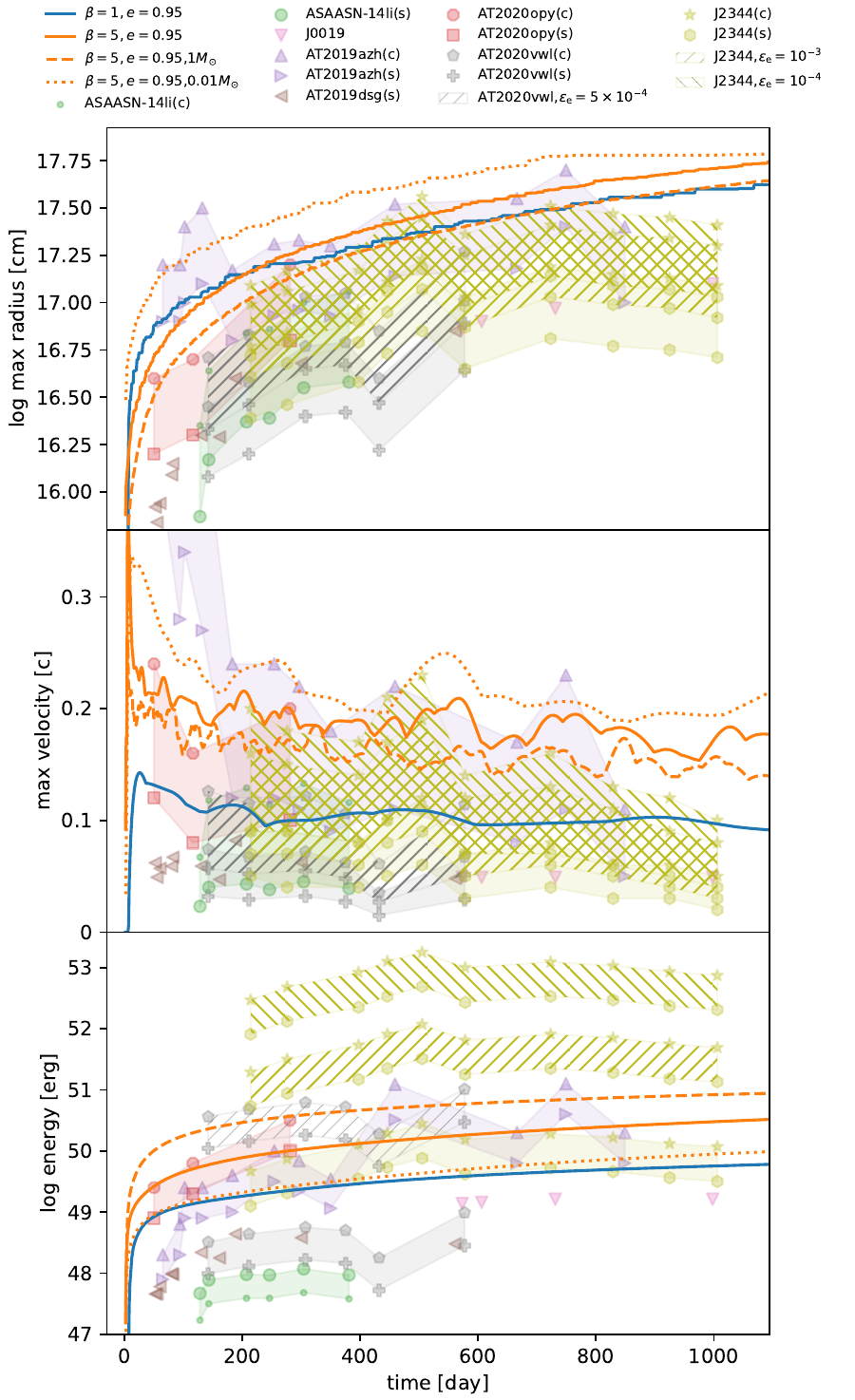}
    \caption{Maximum shocked particle radius (top panel), maximum velocity (middle panel) and total energy (bottom) of the shock region formed between TDE outflow and CNM shell ($M_{\rm CNM} = 0.1M_{\odot}$) with various initial $\beta$ and $e$ (solid lines). Various CNM shell masses are used for the $\beta=5$ TDE (orange lines). Observations are plotted with dots and shaded regions. Shock properties data are available on Zenodo: doi: 10.5281/zenodo.14286338. }
    \label{fig:shock}

\end{figure}

\begin{figure*}
    \centering
    \includegraphics[width=2\columnwidth]{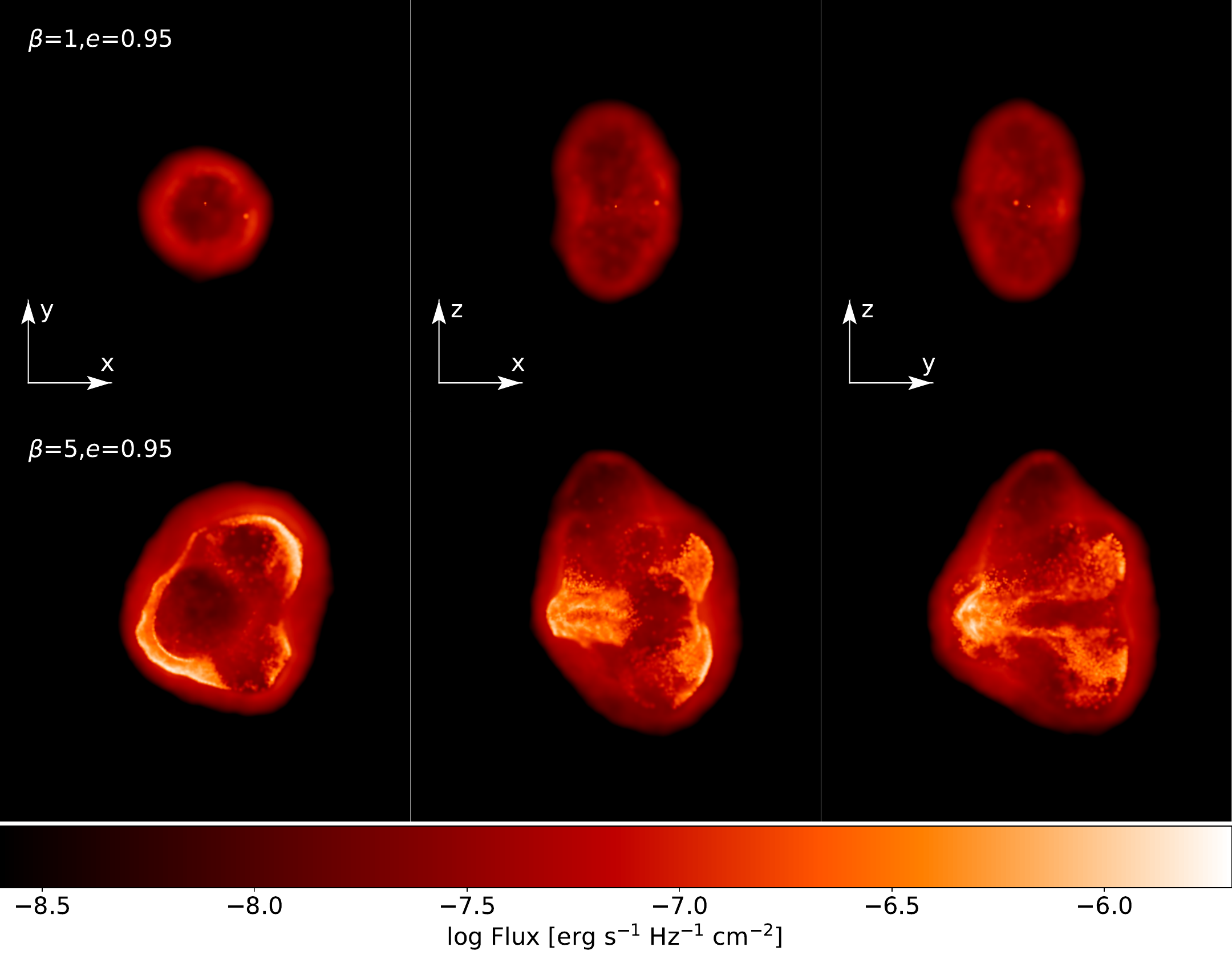}
    \caption{Synthetic images at 5.2 GHz, assuming $\epsilon_{\rm e} = 0.1$ and $\epsilon_{\rm B} = 0.01$, viewed along the z- (left column), y- (middle column) and x-axis (right column), of the synchrotron emission from the outflowing shock shells between various TDEs (top row: $\beta=1,e=0.95$; bottom row: $\beta=5,e=0.95$) and CNM cloud of 0.1~$M_{\odot}$ at $t=3~$yr post first pericenter passage. Each panel is 0.4 pc $\times$ 0.4 pc in size, divided into 1024 $\times$ 1024 pixels, ray traced in each pixel with the corresponding source function and opacity according to Section~\ref{sec:synchrotron spectra}. FITS files resulting from our ray tracing procedure are available on Zenodo: doi: 10.5281/zenodo.14286338.}
    \label{fig:syn img}
\end{figure*}

\subsection{Shock profile} \label{sec:shock profile}
Figure~\ref{fig:shock_profile} shows radial profiles of entropy per baryon (top panel), gas temperature (middle panel) and radial velocity (bottom panel) at 0.3 year post-pericenter passage in our radio simulation with $\beta = 5$ and $M_{\rm CNM} = 0.1 M_{\odot}$. 

The forward shock is moving outward at a speed of $\sim 0.15$c, which gives the pre-shocked CNM a Mach number $\gtrsim 10^4$ relative to the shock front, meaning that a strong shock is formed. The shock dissipates the kinetic energy into thermal/internal energy, raising both the entropy ($\sim 60\%$) and temperature (to $\sim 10^{10}$ K) of the gas.

As shown in the top panel, the pre-shock CNM shell has a variation of $\sim 10\%$ in entropy. If we choose a shock detection threshold of 10\% we would mistakenly detect unshocked particles as shocked. On the other hand, the shock raises the entropy by $\sim 60\%$, so if we choose such threshold we would miss shocked particles. Within the range of $10\%$ to $60\%$, our results do not critically depend on the detection threshold.

\subsection{Shock properties}

In Figure~\ref{fig:shock}, we show the maximum radii, maximum velocities and total energies 
of the shocked particles of all of our simulations with various $\beta$ and $M_{\rm CNM}$, together with the radio observationally-inferred properties of six TDEs (Table~\ref{tab:obs}). In these observations, various outflow geometries and energy fractions are assumed in applying the equipartition model (Section~\ref{sec:equipartition model}) as listed in Table~\ref{tab:obs}. 

Our estimated radius evolution (first panel) are all similar within 0.5 dex, despite varying penetration factor and broadly agree with the observations. TDEs where the pericenter passage is closer to the BH (higher $\beta$) reach larger radii due to the higher outflowing velocity (second panel). Stars passing closer to the BH lead to a stronger self-collision thus more orbital energy dissipated and more mass accreted. This produces more energy to power the outflow and produces faster outflows. All simulations show a peak in the maximum velocity, remaining roughly constant for the rest of the simulation time. The initial peak is due to the earliest outflow being the fastest but slows down rapidly since it contains very little mass \citep[see Figure~3 of][]{Hu2024}. The magnitude of the plateau velocities are strongly related to $\beta$ which increases by a factor of $\sim 2$ from $\beta = 1$ to 5.

The shock radius and velocity is also related to the mass in the circumnuclear medium (orange lines). Lower $M_{\rm CNM}$ corresponds to a lower density in CNM which leads to a faster sound speed and more rapid increase in radius as it takes longer for the shock to decelerate. The plateau in radius for the simulation with $M_{\rm CNM} = 0.01M_{\odot}$ is due to the shock front reaching the outer radius of the CNM shell, i.e. $6\times 10^{17}$ cm. 
 

The total energy (third panel) increases slowly over the three years after a rapid initial rise, as more gas is ejected into outflows from interactions near the SMBH. The initial increase is due to rapid increase in shocked mass according to our shock detection routine. The total energy shares a similar trend to radius and velocity which increases with higher initial $\beta$. 

The energy increases by $\sim$ one order of magnitude from $\beta=1$ to $\beta=5$ due to stronger interaction and greater acceleration near the BH. { The total injected energy is $\sim 7\times10^{50}$ erg and $\sim 3\times10^{51}$ erg for $\beta = 1$ and 5 respectively, of which $\sim $ 3.5\% is in the form of internal energy for both simulations. Meanwhile, the CNM shock energy is about 10\% of the total injected energy with the fraction of internal energy increased to $\sim 50\%$. Therefore, unlike the usual assumption that radio emission energy is of the same order of magnitude as the outflow kinetic energy \citep[e.g.][]{Lu2020,Huang2024c,Zhuang2024}, we found the actual available energy in the shocked region is only $\sim 10\%$ of the total outflow energy.}

$M_{\rm CNM}$ also increases the total kinetic energy due to the increased mass within the shocked region. We see a degeneracy between $\beta$ and $M_{\rm CNM}$, i.e. $\beta=5$ and $M_{\rm CNM}=0.01M_{\odot}$ vs. $\beta=1$ and $M_{\rm CNM}=0.1M_{\odot}$, which might lead to difficulties in observational analysis. 

Our simulations, however, show a narrow range in energy ($\sim$ two order of magnitude) whereas observations show a broader range ($\gtrsim$ four orders of magnitude). This might indicate that $M_{\rm CNM}$ have a greater range than two orders of magnitude in reality, or due to the fixed stellar and BH masses in our simulations, compared to observed TDE outflows which probe a range of stars and black holes.

\subsection{Synthetic images}
In Figure~\ref{fig:syn img}, we show a grid of synthetic images at 5.2 GHz, showing the synchrotron emission of all of our models with $M_{\rm CNM} = 0.1 M_{\odot}$ at 3 yrs post first pericenter passage, viewed from three different viewing angles (same as in Figure~\ref{fig:shock_view}), assuming $\epsilon_{\rm e} = 0.1$ and $\epsilon_{\rm B} =0.1$.

The synchrotron emission mainly comes from the high temperature shell. However the dominant emission region does not coincide with the hottest region. This shows temperature, or the ionisation fraction, is not the dominant factor leading to synchrotron emission, the density is more important. 

From the middle and right columns, it can be seen that the dominant emission regions are symmetric with respect to the horizontal mid-plane which is the initial stellar orbital plane. This shows that most mass is still concentrated near the mid-plane and the outflow possesses more symmetry than previously seen in Figure~\ref{fig:shock_view} or \citet{Hu2024}. Unlike optical emission, the radio emission depends strongly on the density but not temperature which provides a way to study the structure of TDE outflows that is not achievable in optical due to the high opacity of low density yet hot gas.

The synchrotron calculation is not critically resolution dependent (100k for $\beta=1$ simulation in the top row and 1M for $\beta=5$ simulation in the bottom row). 
The radio simulations, however, are mildly resolution dependent.
We perform a resolution study with the $\beta=5,e=0.95$ model to check the effect of numerical resolution (see Appendix~\ref{app:res}).

\subsection{Synthetic spectra}
\begin{figure*}
    \centering
    \includegraphics[height=8.4cm]{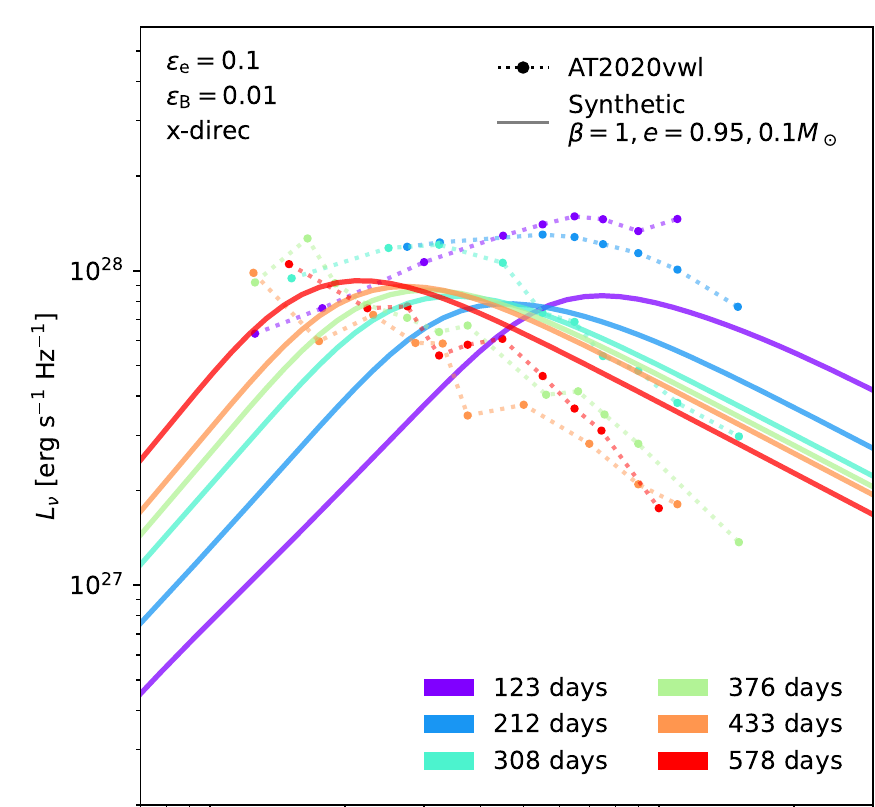}
    \includegraphics[height=8.4cm]{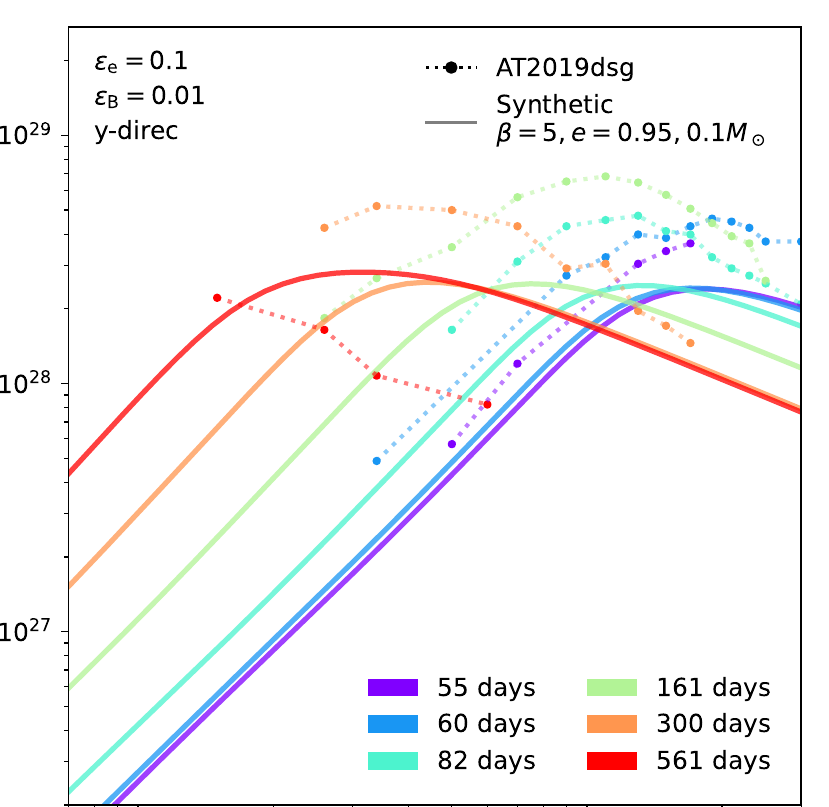}
    \includegraphics[height=9.11cm]{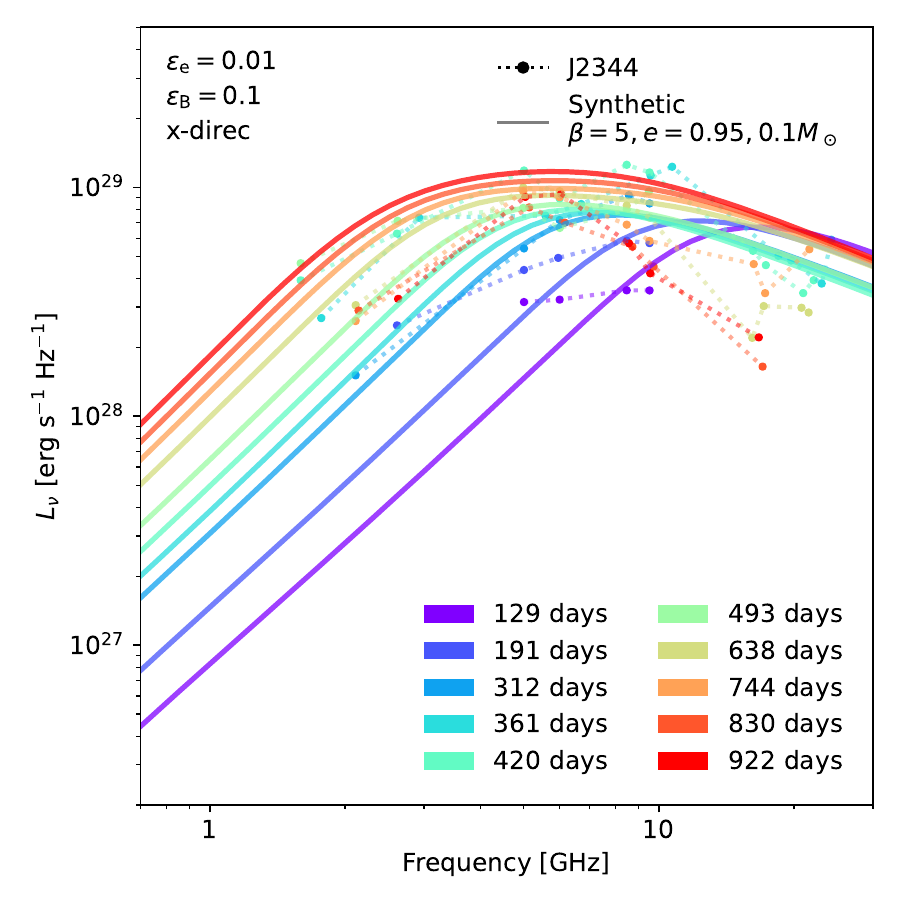}
    \includegraphics[height=9.11cm]{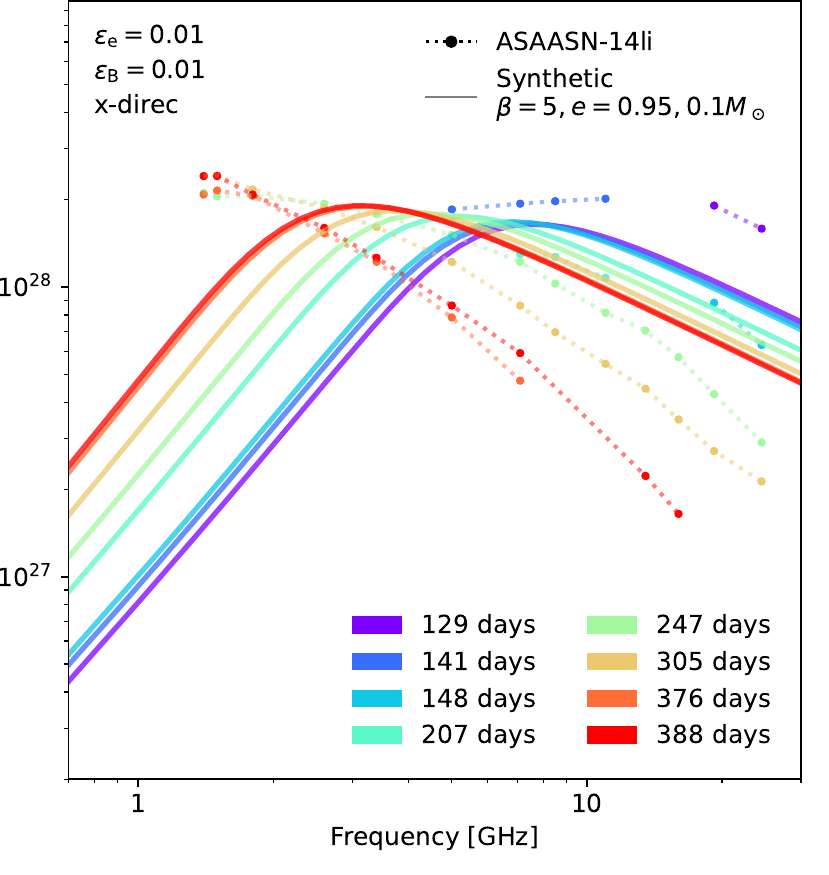}
    \caption{Best fit synthetic spectra (solid lines) to observed radio TDEs (Table~\ref{tab:obs}). Time of the observed spectra (different colours) is from the expected initiation of the optical flare, and time of the synthetic spectra is from the first pericenter passage. Our model TDE with $\beta=5$ and CNM cloud of 0.1 $M_{\odot}$ viewing from the x-direction (i.e. right panel of the second row of Figure~\ref{fig:syn img}) fit the best to 5 out of 7 observed TDEs. AT2020vwl (top left panel) and AT2019dsg (top right panel) are best fitted with matching spectral shape and $\lesssim$ 0.3 dex differences in flux. eRASSt J2344 (bottom left panel) and ASAASN-14li (bottom right panel) have peak frequencies and flux within with same order of magnitude while the spectral shapes do not match as good. For AT2020opy, AT2019azh and CNSS J0019 we do not found reasonable fit. All the synthetic spectra data are available on Zenodo: doi: 10.5281/zenodo.14286338.
    }
    \label{fig:syn spec}
\end{figure*}

In Figure~\ref{fig:syn spec}, we plot the synthetic spectra (solid lines) that best fit to the observed spectra (dotted lines) of AT2020vwl \citep[top left panel;][]{Goodwin2022}, AT2019dsg \citep[top right panel;][]{Cendes2021}, eRASSt J2344 \citep[bottom left panel;][]{Goodwin2024} and ASAASN-14li \citep[bottom right panel;][]{Alexander2016}. We match the time of the observed spectra (time from the initial detection of the optical flare; different colours) with our synthetic ones (time from the first pericenter passage). 

Our synthetic spectra can reproduce the decay in peak frequency whereas the peak flux does not evolve as much as seen in the observed spectra. For AT2020vwl and AT2019dsg, our synthetic spectra match the overall spectral shape of the early observations, i.e. $\lesssim 300$ days. 
After $\sim 300$ days, the evolution of our synthetic spectra is evidently slower than the observed one which could be due to our assumption of a smooth spherically symmetric CNM or the lack of cooling, which becomes important at later times, in our simulations (see Section~\ref{sec:discussion}). For eRASSt J2344 and ASAASN-14li, the peak frequencies of our synthetic spectra are within the same order of magnitude of the observations but do not match as good as AT2020vwl and AT2019dsg.

The peak flux of all of our synthetic spectra only evolve by a factor of $\lesssim 2$ over the three year period, whereas the observations show a large variation, e.g. AT2019dsg.
In our simulations the main parameter that affect the flux is the magnetic field, which is related to $\epsilon_{B}$ and density. Therefore a clumpy CNM or CNM with steeper density profiles could lead to the greater variation in peak flux (see Section~\ref{sec:discussion}). On the other hand, we do not vary the stellar mass or BH mass in the work, which are also likely to affect the flux.

\subsection{Peak frequency}
\begin{figure}
    \centering
    \includegraphics[width=\linewidth]{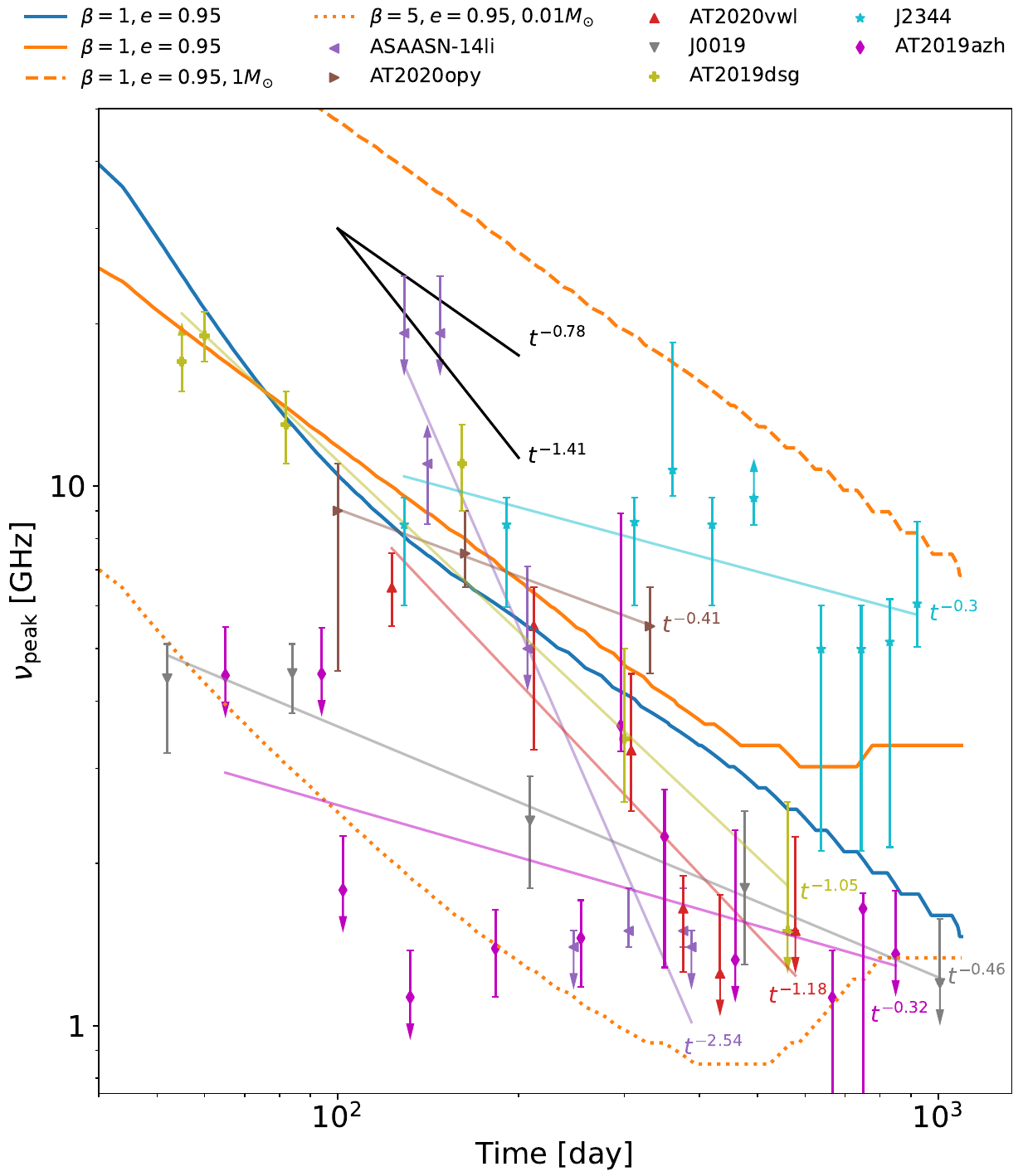}
    \caption{Temporal evolution of peak frequency, i.e. the frequency of peak flux, assuming $\epsilon_{\rm e} = 0.1$ and $\epsilon_{\rm B} = 0.01$, of the synchrotron spectra. The synthetic peak frequencies (lines) show slight variation in decay rates (black lines), i.e. early time of the model with $\beta=1$ decays as $t^{-1.41}$ (blue line at $t\lesssim 100$ days), the ones with $\beta=5$ and late time of the one with $\beta=1$ decay as $t^{-0.78}$ (orange lines; blue line at $t \gtrsim 100$ days). The observed peak frequencies (dots) show a larger variety of evolution trend (opaque lines). Arrows mark the data points being a lower or upper limit. Our synthetic models are within the range of the observations, i.e. $t^{-0.3}$ to $t^{-2.54}$.}
    \label{fig:peak freq}
\end{figure}

In Figure~\ref{fig:peak freq}, we plotted the evolution of peak frequency against time of our synthetic spectra (lines) and the observations (dots). When the peak frequency occur on the boundary of the observing range, it is a lower or upper limit, which we marked with arrows in Figure~\ref{fig:peak freq}.
We fit the line of best fit to each event and plotted with opaque lines. Our synthetic spectra show the same temporal evolution: peak frequency decays with $t^{-0.78}$ in the one with $\beta = 5$ and late time ($\gtrsim 100$ days) of the one with $\beta = 1$, whereas the observations vary from $t^{-0.3}$ \citep[eRASSt J2344;][]{Goodwin2024} to $t^{-2.54}$ \citep[ASAASN-14li;][]{Alexander2016}. Given the inhomogeneity of the CNM, a single power law cannot fully characterise the observed data, but it roughly matches the evolutional trend.

The CNM mass, or density, does not affect the evolution power law but only the peak frequency. An increase by a factor of 10 in the CNM mass corresponds to an increase by a factor of $\sim 4$ in the peak frequency.

In the models with $\beta = 5$, a rise in peak frequency appear after $\sim 40$ days for CNM mass of $0.1M\odot$ and $0.01M\odot$. This is due to the differences in opacity between CNM shells. The emission from early shocked mass fades as electron cools with time, i.e. the time dependence of cooling frequency (Equation~\ref{eq:nu c}), and the later shocked mass is still emitting. As the opacity drops as the gas evolves, it could become low enough to allow the emission from the later shocked mass to escape and a rise in the peak frequency is observed.

\section{Discussion} \label{sec:discussion}


{
We determine which particles from the TDE simulation to be injected into the radio one exclusively based on distance reached, i.e. all particles reaching $r_{\rm inj}$ in the TDE simulation are injected into the radio one, whether bound ($E<0$) or unbound ($E>0$). We found that among all injected particles, 15\% (35\%) are still bound to the BH, which reduces the injected energy by 3.6\% (10\%) for the $\beta=5$ (1) simulation. Therefore, the bound gas has a minor impact on the interaction with CNM, but is unlikely to qualitatively change the results.
}

The interaction between the TDE outflow and the CNM is similar to supernova remnants which also form from the collision between the outflowing gas and the interstellar medium \citep[e.g.][]{Mandal2023}. It is possible to explain the radio emission, $\gamma$-ray emission and neutrinos observed from TDEs \citep[e.g.][]{Murase2020, Goodwin2023} by such interactions, similarly to supernova remnants \citep[e.g.][]{Fukui2024}. Both of the maximum radii and maximum velocities of our shocks are of the similar order of magnitude as the observational data, indicating the radio emission is consistent with the interactions between the TDE outflow and the CNM with density $\sim 10^{-17}$ g cm$^{-3}$. In our models, the maximum density of the CNM is comparable with the minimum density of the TDE outflow. Given the outflow would depend on the initial stellar mass and BH mass, the density of CNM to produce observed radio emission could vary.

To infer the properties from observations, one needs to assume the proportion of energy carried by electrons, $\epsilon_{\rm e}$, which is not a well constrained parameter and commonly assumed to be $\epsilon_{\rm e} = 0.1$ (see Table~\ref{tab:obs}). Recent investigations \citep[e.g.][]{Park2015, Xu2020}, however, suggest that electrons are less effectively accelerated than protons in such non-relativistic quasi-perpendicular collisionless shocks which results in less energy carried by electrons. \citet{Goodwin2023} and \citet{Goodwin2024} have attempted to fit with lower $\epsilon_{\rm e}$. In Figure~\ref{fig:shock}, we show the inferred properties with $\epsilon_{\rm e} = 5\times 10^{-4}$ (cyan left hatch) for AT2020vwl, $10^{-3}$ (magenta left hatch) and $10^{-4}$ (magenta right hatch) for eRASSt J2344. The choice of $\epsilon_{\rm e}$ does not have a significant effect on the inferred radii and velocities, but causes changes by a few orders of magnitude in the inferred energy. 
Compared to our estimates, most of the observations have lower total energy when assuming $\epsilon_{\rm e} = 0.1$, however, the inferred energies would become comparable or even higher with lower $\epsilon_{\rm e}$. 

The geometry of the TDE outflow is also not well constrained. In observations, spherical (s; $f_{\rm A} = 1$) and conical (c; $f_{\rm A} \approx 0.1$) geometries are usually assumed to capture the uncertainties (see Table~\ref{tab:obs}). An intermediate geometry is also used by \citet{Alexander2016} with $f_{\rm A} = 1/3$. The spherical geometry assumes the synchrotron emission is powered by TDE outflow whereas the conical geometry assumes it is powered by jets. As seen in Figure~\ref{fig:shock_view} and \ref{fig:syn img}, our outflows or the shock regions are quasi-ellipsoidal and the regions that dominate the emission are closer to a doughnut shape near the initial orbital plane (see the second row of Figure~\ref{fig:syn img}). The emitting region geometry however depends on the penetration factor and likely the initial orbital eccentricity. Further investigations are necessary to better justify the choice of models for interpreting observations.


By comparing our synthetic spectra with the observations, we find that we need $\epsilon_{\rm e} \sim 0.1$ and $\epsilon_{\rm B} \sim 0.01$ to produce spectra that are close to the observed ones in both luminosity and shape (top row of Figure~\ref{fig:syn spec}). This is consistent with the result of \citet{Cendes2021} from the cooling break, i.e. $\epsilon_{\rm B} \approx 0.02$, as well as the general assumptions in observation, i.e. $\epsilon_{\rm e} = 0.1$. The offset is possibly due to the same reasons as why we see no decrease in peak flux (see later paragraphs). Within the parameter space we simulated, we see that $\epsilon_{\rm B}$ is the main parameter that affect the flux, whereas $\epsilon_{\rm e}$ has little effect when smaller than $\sim 0.1$. 

We see that the observations are generally steeper than our synthetic spectra at frequencies higher than the peak frequency. This indicates in TDEs the electron distribution power law index $p$ is higher than the common assumption for GRBs, i.e. $\sim 3$ instead of 2.5. On the other hand, a CNM shell with lower or higher density tends to shift the peak towards lower or higher frequency respectively, due to the change in opacity. In our models with spherically symmetric CNM shell, the viewing angle is not critical to the resulting spectra. In a real CNM shell, however, due to the inhomogeneity we expect that viewing angle would be important.


The observed spectra continuously decreases in peak frequency while the peak flux raises and decays over time. Our synthetic spectra does not show as strong variation in peak flux as the observations and the decay rate of peak frequency is slower than the observations. Possible explanations include:

\begin{enumerate}
    \item The CNM density profile is steeper or the CNM shell is thinner. The continuous increase in flux could be due to the continuous shock formation, which keeps adding new sources of synchrotron emission. 
    With a steeper density profile or a cut off in CNM shell, shock formation reduces more rapidly or even stops at later time. Given the synchrotron emission decay with time as electrons cooled and recombined, without enough new sources, the overall spectra could decay.
    \item We did not account for cooling properly (e.g. energy lost through synchrotron, thermal and bremsstrahlung emission), which leads to an overestimation of the number of the emitting electrons at late times. In our models the emitting region expands with the outflow but the flux decays at a slower rate so the luminosity continues growing.
    \item The CNM is not smooth. If denser clumps are present in the CNM shell, bow shocks will form when the outflow interacts with the clouds \citep[see e.g.][]{Zhuang2024}. This disturbs the outflow and the CNM which causes more interactions and hastens evolution.
    \item The energy partition changes as the TDE evolves, i.e. $\epsilon_{\rm B}$ drops over time. 
    The observed spectra are consistent with our synthetic ones in flux at later times when $\epsilon_{\rm B}$ drops by $\sim$ one order of magnitude.
\end{enumerate}
Further investigations are necessary to investigate which of these hypotheses are correct.

The variance in the evolution power law is mainly due to the variance in the structure of the CNM shell. The initial penetration factor and eccentricity only have small effects. Compared with AT2019dsg before $\sim 200$ days \citep{Cendes2021}, where we adopted the CNM density power law index from, the evolution of the peak frequency is roughly consistent with our models. Future investigations are necessary to understand the connections between peak frequency evolution and CNM structures.

{
Our synthetic spectra (Figure~5) show that the peak frequency $\nu_{\rm peak}$ does not coincide with either of the local break frequencies, $\nu_m$ or $\nu_c$, but is instead primarily determined by the self-absorption frequency $\nu_a$. This arises because the minimum Lorentz factor of shock-accelerated electrons remains close to $\gamma_m \simeq 1$, causing $\nu_m$ to fall into the MHz regime, while the observed GHz peak emerges near the frequency where the optical depth becomes unity. The resulting spectral shape is consistent with Spectrum~2 shown in Figure~\ref{fig:benchmark}.
s
The scaling of $\nu_a$ can be expressed by Equation~\eqref{eq:nua} as
\begin{equation}
\nu_a \propto \rho^{\frac{2}{p+4}} B^{\frac{p+2}{p+4}} R^{\frac{2}{p+4}},
\label{eq:nuascale}
\end{equation}
where $\gamma\approx1$ was adopted. Assuming $B \propto R^{-m}$, Equation~(\ref{eq:nuascale}) becomes
\begin{equation}
\nu_a \propto R^q, \quad q \equiv \frac{2}{p+4}n - \frac{p+2}{p+4}m + \frac{2}{p+4},
\nonumber
\end{equation}
where we used Equation~(\ref{eq:rho}) for the derivation. Combining with the radial expansion $R \propto t^\delta$ yields the temporal evolution of the peak frequency as
\begin{equation}
\nu_{\rm peak} \simeq \nu_a \propto t^{q\delta}.
\nonumber
\end{equation}
For representative parameters ($n = -1.7$, $m = 1$, $p = 2.5$), we find $q \simeq -0.91$ and, with the simulated $\delta \simeq 0.78$ (Figure~\ref{fig:shock}), this gives $q\delta \simeq -0.71$, in agreement with the decay slope of $\nu_{\rm peak} \propto t^{-0.78}$ shown in Figure~\ref{fig:peak freq}.

Figure~\ref{fig:peak freq} also shows that the observed evolution of $\nu_{\rm peak}$ spans a broad range from $t^{-0.3}$ to $t^{-2.5}$.
Equation~(3) suggests that this observed diversity can be explained primarily by variations in the expansion index $\delta$. In particular, $\delta$ depends on the ambient density slope $n$ through the power-law solutions:
\begin{equation}
\delta_{\rm ST} = \frac{2}{5+n}, \quad \delta_{\rm SP} = \frac{1}{4+n},
\nonumber 
\end{equation}
corresponding to Sedov–Taylor phase \citep{Sedov1959} and momentum-conserving snow-plow phase in the context of TDE outflows \citep{Hayasaki2023}, respectively. For $n = -1.7$, we obtain $\delta_{\rm ST} \simeq 0.61$ and $\delta_{\rm SP} \simeq 0.43$. Combined with $q \simeq -0.91$, this yields $q\delta_{\rm ST} \simeq -0.56$ and $q\delta_{\rm SP} \simeq -0.39$, which are consistent with the more gradual decay slopes observed in some events. Steeper declines such as $\nu_{\rm peak} \propto t^{-2.54}$ may require larger $\delta$ values, possibly reflecting an earlier free-expansion phase, or alternatively flatter ambient density profiles ($n \rightarrow 0$) or more rapid magnetic field decay ($m > 1$).

In summary, identifying $\nu_{\rm peak} \simeq \nu_a$ and applying the corresponding scaling provides a physically consistent framework to interpret our simulation results, analytic models (e.g., the Sedov-Taylor or snow-plow phases), and the diverse temporal behavior seen in observations. Further applications to a broader range of initial conditions and comparison with observational data will help assess the robustness of this framework.
}

\section{Conclusions}
We set up a spherically symmetric circumnuclear material shell and inject a TDE outflow to study their interaction. Shocks formed at the interface as expected. The behaviour of the shocked regions is similar:
\begin{enumerate}
    \item Maximum radius increases linearly.
    \item The maximum velocity peaks initially and then remains roughly constant for three years.
    \item Total energy increases rapidly at $t \lesssim 100$ days up to $\sim 10^{49}$--$10^{51}$ erg and slowly increase by factors of a few over three years.
    \item { We found the total energy available for synchrotron radiation to be $\sim 10\%$ of the total outflow energy, about one order of magnitude lower than previous assumptions. }
\end{enumerate}
The radius and velocity shows the same order of magnitude and evolution as observations whereas the total energy of our simulations has a smaller range compared to observations.

We then ray traced the synchrotron emissions from the shocked regions to produce synthetic images and spectra and compare with the observations. We found that:
\begin{enumerate}
    \item An electron energy fraction of $0.1$ and magnetic energy fraction of $0.01$ produce spectra that best match the observations over a broad range of the penetration factor and eccentricity.
    \item Our synthetic spectra show continuous decay in peak frequency similar to the early stage of the observation. The decay rate is slower than the observations at late time.
    \item Our synthetic spectra do not reproduce the variations in peak flux in observations.
    \item The evolution of the peak frequency of our synthetic spectra is within the range of the observed rates.
\end{enumerate}

Our models support the hypothesis that the synchrotron radio flares are from the collision between the TDE outflows and circumnuclear material (hypothesis 2). 
Given that outflows are a common feature of TDEs, whereas jets are  rare among TDEs, we expect outflows to be the dominant source of radio synchrotron emission from TDEs. TDEs with circumnuclear material dense enough to form significant collision with the outflow could produce observed synchrotron radio emission.

\begin{acknowledgments}
This research was supported in part by grant NSF PHY-2309135 to the Kavli Institute for Theoretical Physics (KITP), where most of us were hosted for the tde24 workshop in April--May 2024. We acknowledge the CPU time on OzSTAR and Ngarrgu Tindebeek funded by the Victorian and Australian governments and Swinburne University. F.H. acknowledges support from Astronomical Society of Australia (ASA), through ASA Student Travel Grant. F.H. \& I.M. acknowledge support from the Australian Research Council Centre of Excellence for Gravitational Wave Discovery (OzGrav), through project No. CE230100016. K.H. has been supported by the Basic Science Research Program through the National Research Foundation of Korea (NRF) funded by the Ministry of Education (2016R1A5A1013277 and 2020R1A2C1007219 to K.H.).

\end{acknowledgments}

\bibliography{ref.bib}{}
\bibliographystyle{aasjournal}

\renewcommand\thefigure{\thesection.\arabic{figure}}   
\setcounter{figure}{0}    

\appendix
\section{Resolution study} \label{app:res}
To study the effects of resolution on the radio simulations and the synchrotron ray tracing, we run three simulations with $\beta=5, e=0.95$ and resolutions of 100k, 1M and 10M SPH particles. 

\begin{figure}
    \centering
    \includegraphics[width=\linewidth]{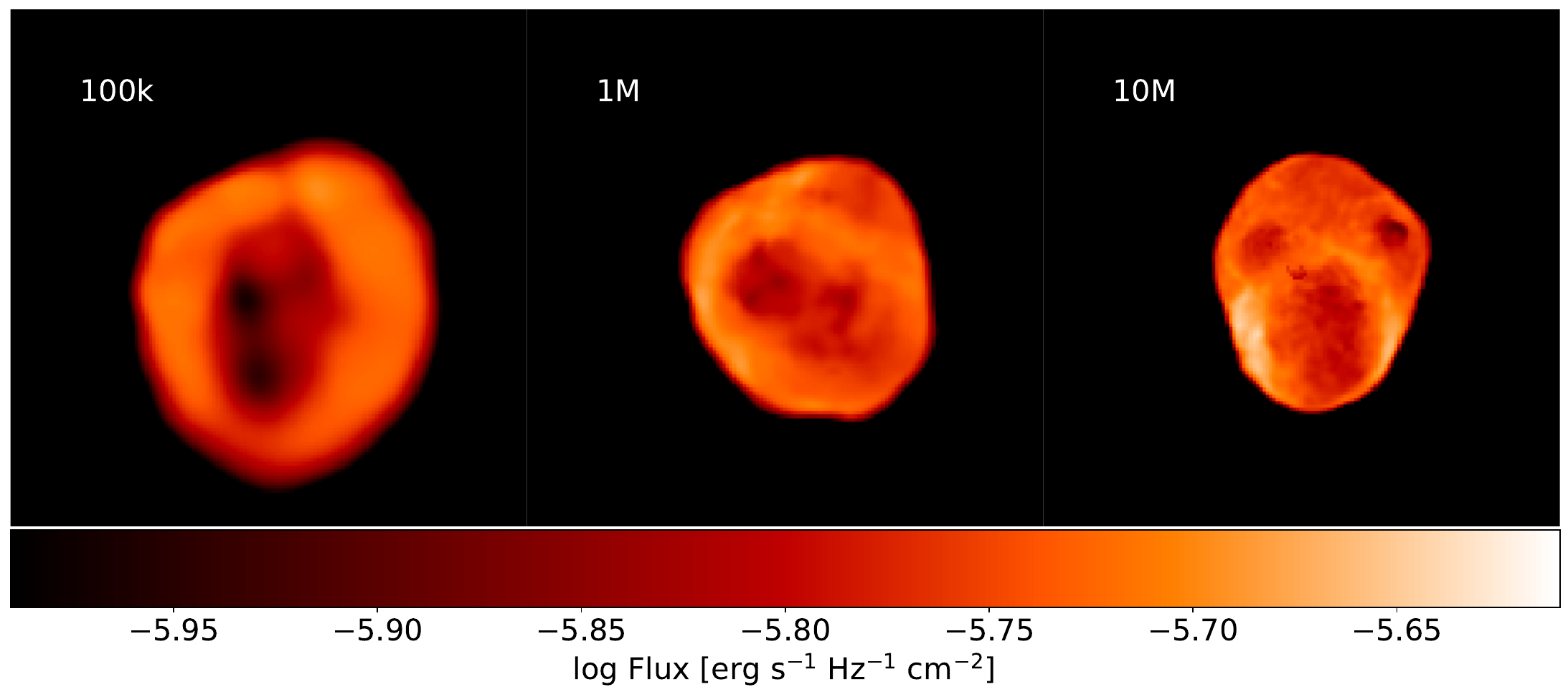}
    \caption{Synthetic images at 5.2 GHz, assuming $\epsilon_{\rm e} = 0.1$ and $\epsilon_{\rm B} = 0.01$, viewed along the z-axis, of the synchrotron emission from the outflowing shock shells between TDEs of $\beta=5,e=0.95$ and resolution of 100k (left), 1M (middle), 10M (right), and CNM cloud of 0.1 $M_{\odot}$ at $t=0.3~$yr post first pericenter passage. Each panel is 1024 $\times$ 1024 pixels, ray traced in each pixel with the corresponding source function and opacity according to Section~\ref{sec:synchrotron spectra}. The emitting area shrinks with resolutions whereas the flux is converged.}
    \label{fig:res img}
\end{figure}

In Figure~\ref{fig:res img}, we plot the synchrotron images at 5.2 GHz, assuming $\epsilon_{\rm e} = 0.1$ and $\epsilon_{B} = 0.01$, for the 100k (left), 1M (middle) and 10M (right) simulations, viewed along the z-direction (similar to the left column of Figure~\ref{fig:syn img}) but at $t = 0.3$ yr. The emitting area shrinks with resolutions due to a slower shock wave speed in the higher resolution CNM shell, whereas the flux is converged.

\begin{figure}
    \centering
    \includegraphics[width=0.9\linewidth]{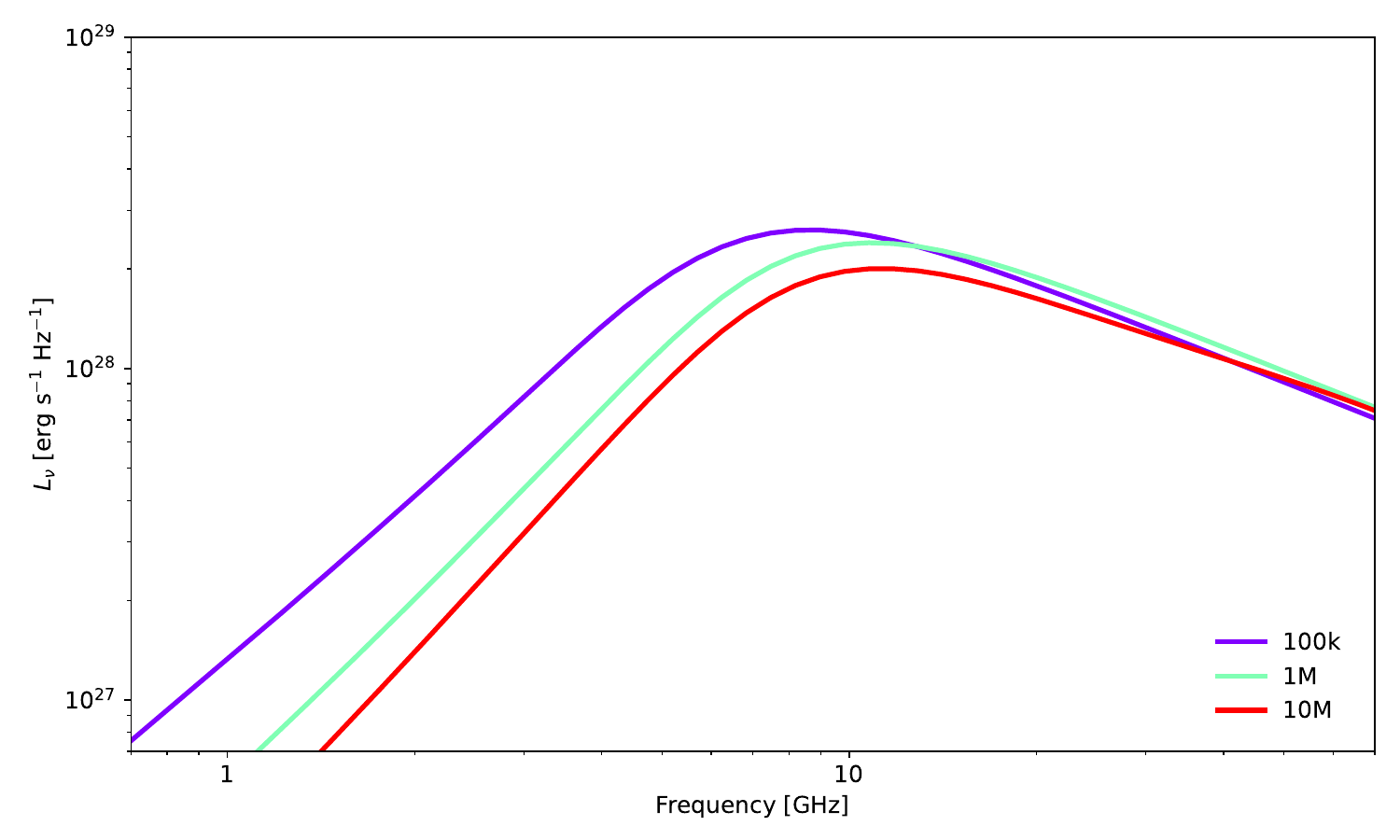}
    \caption{Synthetic spectra, assuming $\epsilon_{\rm e} = 0.1$ and $\epsilon_{\rm B} = 0.01$, of the synchrotron emission from the outflowing shock shells between the model TDE with $\beta=5$ and resolution of 100k (purple), 1M (green), 10M (red), and CNM cloud of 0.1 $M_{\odot}$, i.e. Figure~\ref{fig:res img}. Peak frequency and peak flux change by less than a factor of two between various resolutions. Flux above $\sim 20$ GHz is converged, whereas below $\sim 20$ is not.}
    \label{fig:res spec}
\end{figure}

In Figure~\ref{fig:res spec}, we also plotted the spectra of the images in Figure~\ref{fig:res img}. The peak frequency and peak flux change by less than a factor of two when the resolution increases by a factor of 100. The flux above $\sim 20$ GHz is converged whereas below $\sim 20$ GHz it is not yet converged. The critical parts we are interested are the peak frequency and peak flux which are not strongly resolution dependent.

\begin{figure}
    \centering
    \includegraphics[width=0.8\linewidth]{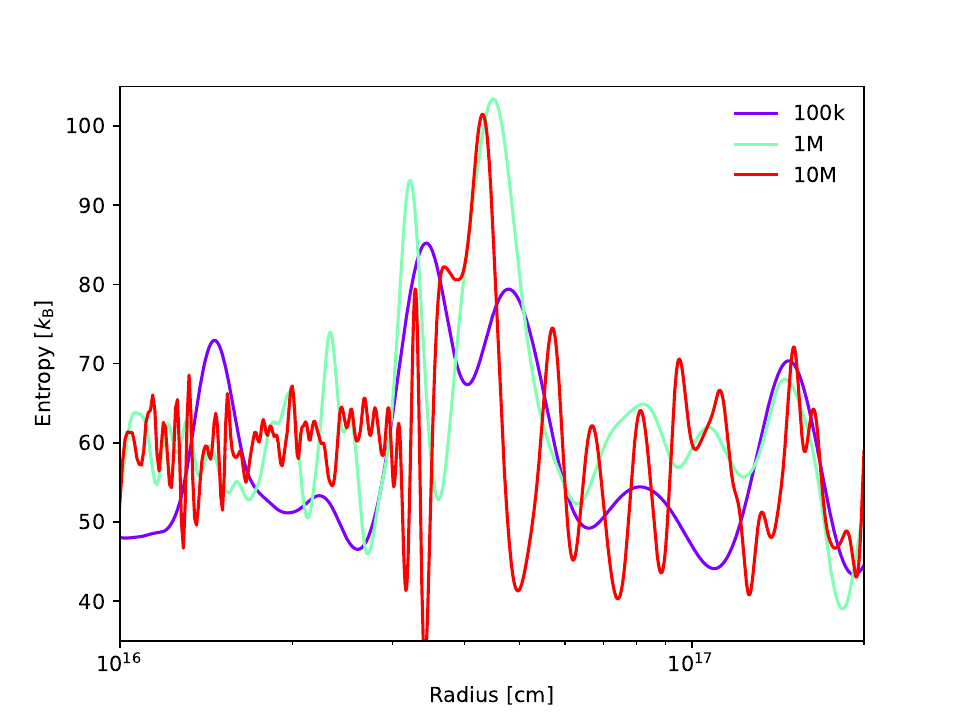}
    \caption{Entropy profile along a single line of sight, i.e. positive x-axis in Figure~\ref{fig:res img} and \ref{fig:syn img} at resolution of 100k (purple), 1M (green), 10M (red). The max entropy is the same between 1M and 10M ones and the shock width decreases with resolution due to smaller spacing between particles. The fluctuations are due to the interpolation of SPH particles properties onto the line of sight \citep[for more information see][]{Price2007}.} 
    \label{fig:shock res}
\end{figure}

In Figure~\ref{fig:shock res}, we plotted the entropy profile along a single line of sight, i.e. positive x-axis in Figure~\ref{fig:res img} and \ref{fig:syn img}, at different resolutions. It can be seen that in the shocked region, the entropy increase is the same for 1M and 10M simulations. The width of the shock reduces with resolution. This is purely due to the larger spacing between particles and does not change the shock strength or heat generation. The fluctuations are due to the interpolation of SPH particles properties onto the line of sight \citep[for more information see][]{Price2007}.

\section{Benchmark} \label{app:benckmark}
To verify our synchrotron ray tracing scheme, we simulate a homogeneous cube of gas, assuming pure and fully ionised hydrogen, which relates the electron density to gas density with $n_{\rm e} = \rho/m_{\rm p}$.
Given Equation~\eqref{eq:mag strength}, \eqref{eq:nu m} and \eqref{eq:nu c}, we can express $\nu_{\rm m}$ and $\nu_{\rm c}$ as functions of $\gamma$ and $\rho$, i.e.
\begin{align}
    \nu_{\rm m} &= \sqrt{\frac{8}{\pi}} \epsilon_{\rm B}^{1/2} \epsilon_{\rm e}^{2} \left( \frac{p-2}{p-1} \right)^2 \frac{q_{\rm e}m_{\rm p}^2}{m_{\rm e}^3} \gamma^4 \rho^{1/2}, \\
    \nu_{\rm c} &= \frac{9}{128}\sqrt{\frac{2}{\pi}} \epsilon_{\rm B}^{-3/2} \frac{q_{\rm e} m_{\rm e}}{\sigma_{\rm T}^2 c^2} t^{-2} \gamma^{-4} \rho^{-3/2}.
\end{align}
At the absorption frequency $\nu_{\rm a}$, the optical depth exceeds 1, i.e.
\begin{equation*}
    \tau_{\nu_{\rm a}} = \int_s \kappa_{\nu_{\rm a}} \rho \d s = 1.
\end{equation*}
Since the gas is uniform, the optical depth and density is the same everywhere, and this can be simplified to
\begin{equation} \label{eq:benchmark optical thick}
    \kappa_{\nu_{\rm a}} \rho s = 1,
\end{equation}
where $s$ is the size of the gas cube.
Combine Equation~\eqref{eq:abs coe} and \eqref{eq:benchmark optical thick}, we can express the absorption frequency as
\begin{equation}
    \nu_{\rm a} = \left[\frac{\sqrt{6\pi}q_{\rm e}^3 s}{2\pi m_{\rm e}m_{\rm p}^2c} \left(\frac{6\sqrt{2\pi}q_{\rm e}m_{\rm p}^2}{\pi m_{\rm e}^3}\right)^{\frac{p}{2}}  \epsilon_{\rm B}^{\frac{p+2}{4}} \epsilon_{\rm e}^{p-1} \frac{(p-2)^{p-1}}{(p-1)^{p-2}} \Gamma\left(\frac{3p+2}{12}\right)\Gamma\left(\frac{3p+22}{12}\right)\right]^{\frac{2}{p+4}} \gamma^{\frac{3p}{p+4}} \rho^{\frac{p+6}{2p+8}}.
    \label{eq:nua}
\end{equation}

\begin{figure}
    \centering
    \includegraphics[width=0.5\linewidth]{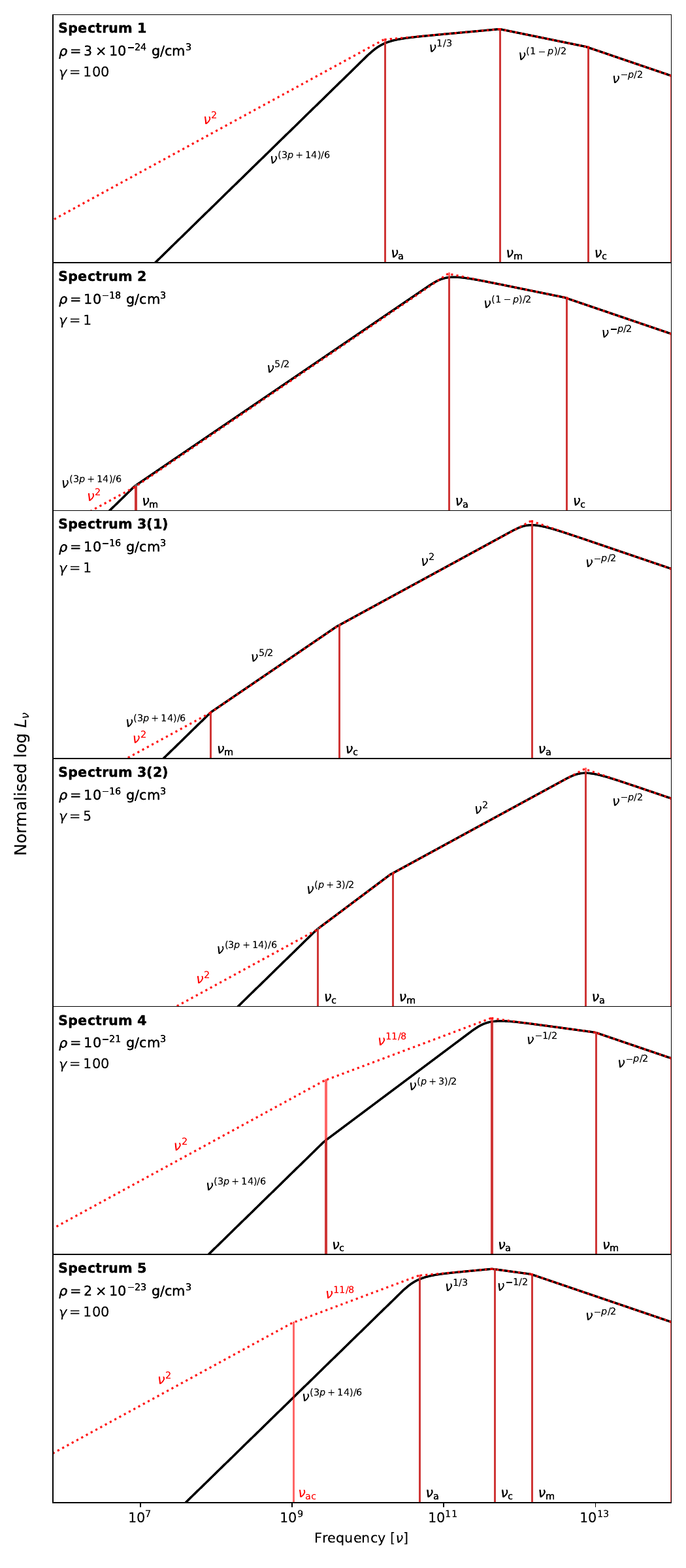}
        \caption{Synthetic spectra of a homogeneous cube of gas of pure and fully ionised hydrogen, with various density and Lorentz factors. Depending on the order of the breaking frequencies $\nu_{\rm m}$, $\nu_{\rm c}$ and $\nu_{\rm a}$, the spectra can be characterised into six types. The label is matched with \citet{Granot2002}, whose solutions are modified and plotted with red dotted lines. The lowest segment has a power law of $\nu^{(3p+14)/6}$ instead of the $\nu^{2}$ Rayleigh-Jean tail since we did not include blackbody emission in our source function. We did not account for a separate source of opacity from uncooled electrons, therefore we we miss the $\nu^{11/8}$ segment in the fast cooling cases.}
    \label{fig:benchmark}
\end{figure}

With our chosen $\epsilon_{\rm B} = 10^{-4}$, $\epsilon_{\rm e} = 10^{-3}$ and $p = 2.5$ we can calculate $\nu_{\rm m}$, $\nu_{\rm c}$ and $\nu_{\rm a}$ from various $\rho$ and $\gamma$, and use certain combinations to produce each spectrum as shown in Figure~\ref{fig:benchmark}. 

Depending on the order of $\nu_{\rm m}$, $\nu_{\rm c}$ and $\nu_{\rm a}$, the spectra can be characterised into six types which we labelled in consistent with \citet{Granot2002}. Spectra 1, 2 and 3(1) correspond to the slow cooling regime which we use the spectrum in Equation~\eqref{eq:spec}. Spectra 3(2), 4, 5 correspond to the fast cooling regime which we instead use the spectrum
\begin{equation} \label{eq:spec fast}
    P_{\nu} = \left\{
    \begin{array}{ll}
        (\nu/\nu_{\rm c})^{1/3} P_{\nu, {\rm max}} & \nu_{\rm c} \geq \nu \\
        (\nu/\nu_{\rm c})^{-1/2} P_{\nu, {\rm max}} & \nu_{\rm m} \geq \nu > \nu_{\rm c} \\
        (\nu_{\rm m}/\nu_{\rm c})^{-1/2}(\nu/\nu_{\rm m})^{-p/2} P_{\nu, {\rm max}} & \nu > \nu_{\rm m}
    \end{array}
    \right..
\end{equation}

There are a few exceptions in the spectral segments compared with \citet{Granot2002}. The lowest segment (segment B in \citealt{Granot2002}) has a power law of $\nu^{(3p+14)/6}$ instead of the $\nu^{2}$ Rayleigh-Jean tail since we did not include blackbody in our source function. We did not account for a separate source of opacity from uncooled electrons, therefore we we miss the $\nu^{11/8}$ segment (segment C in \citealt{Granot2002}) in the fast cooling cases (Spectrum 4 and 5).

We found the order of $\nu_{\rm c}$ and $\nu_{\rm m}$ affects the spectrum which are shown as Spectrum 3(1) and 3(2). We found an extra power law segment of $\nu^2$ between $\nu_{\rm a}$ and the higher of $\nu_{\rm c}$ and $\nu_{\rm c}$. When $\nu_{c} < \nu_{m}$ we found a segment of $\nu^{(p+3)/2}$ instead of $\nu^{5/2}$. We plotted the modified analytical solutions are red dotted lines in Figure~\ref{fig:benchmark}.

\section{Effect of asymmetries}
To study the effect of asymmetries in TDE outflows (which is intrinsic to our TDE simulations), we perform an additional radio simulation with the TDE outflow spherically averaged, which may be directly compared to previous analytic studies by \citet{Matsumoto2021,Matsumoto2024}.

Instead of injecting particles from TDE simulation directly into radio simulation with the original position, velocity, density and internal energy, we average the kinetic and internal energy across all injected particles during each timestep. We then place the injected particles in a spherically symmetric manner as a thin shell with radius $r = (r_{\rm inj} + r_{\rm 0})/2$ and set their radial velocity according to the averaged kinetic energy. The spherically averaged outflow then collides with the spherically symmetric CNM shell (Section~\ref{sec:cnm setup}) as usual. 

\begin{figure}
    \centering
    \includegraphics[width=\linewidth]{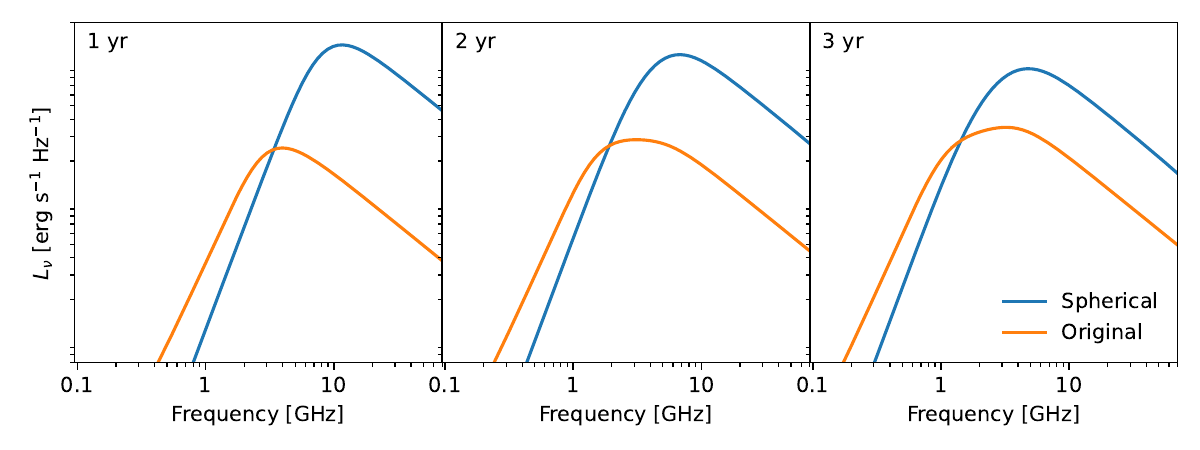}
    \caption{Spectra of synchrotron emission from the collision between the original (asymmetric; orange) and spherically averaged outflow (blue) with the $0.1 M_{\odot}$ CNM (Section~\ref{sec:cnm setup}), at $t = 1$, 2 and 3 years. With the spherically averaged outflow, the peak flux decreases with time instead of increasing, and the spectral shape remain unchanged for the whole period of 3 years. }
    \label{fig:asymmetry}
\end{figure}

In Figure~\ref{fig:asymmetry}, we show the comparisons between the original spectra (orange) and the spherically averaged spectra (blue) for the model with $\beta = 5$ and $M_{\rm CNM} = 0.1M_{\odot}$ at $t =1$, 2 and 3 years. 

It can be seen that the evolution of peak flux is opposite where the original is increasing and the spherical is decreasing. This is because the spherically averaged outflow produces a large spherical emitting region from the initial interaction, whereas the original outflow is concentrated in a small region. Since the earliest outflow has the greatest velocity, later outflow cannot reach earlier one to create new shocks thus the flux keeps dropping due to electron cooling. However with asymmetry, later outflow could collide with CNM in other directions and keep forming shocks to increase total flux.

The shape of the spherical spectra remain roughly unchanged during the entire 3 years whereas the original one flattens. This is due to the collision between CNM and outflow of various velocities produces synchrotron radiation that peaks at various frequencies. As seen in the observations (e.g. Figure~\ref{fig:syn spec}), the spectral shape does change with time, therefore a variety in outflow velocities is necessary for radio TDE analysis.

\end{document}